\documentclass[twocolumn, superscriptaddress ,prc
]{revtex4-2}

\usepackage{graphicx}
\usepackage{dcolumn}
\usepackage{bm}
\usepackage{tabularx}
\usepackage{longtable}
\usepackage{ltablex}
\usepackage{multirow}
\usepackage[T1]{fontenc}
\usepackage[utf8]{inputenc}
\usepackage{color, colortbl}
\definecolor{Gray}{gray}{0.9}
\usepackage[english]{babel}
\usepackage{blindtext}
\usepackage{textgreek}
\usepackage{rotating}
\usepackage{float}
\usepackage{threeparttablex} 
\usepackage{footnote}\makesavenoteenv{longtable}

\begin{document}


\title{Spectroscopic studies of neutron-rich $^{129}$In and its $\beta$-decay daughter, $^{129}$Sn, using the GRIFFIN spectrometer}
\author{F.H.~Garcia}
\email{fatimag@sfu.ca}
\affiliation{Department of Chemistry, Simon Fraser University, Burnaby, British Columbia V5A 1S6, Canada}
\author{C.~Andreoiu}
\affiliation{Department of Chemistry, Simon Fraser University, Burnaby, British Columbia V5A 1S6, Canada}
\author{G.C.~Ball}
\affiliation{TRIUMF, 4004 Wesbrook Mall, Vancouver, British Columbia V6T 2A3, Canada}
\author{N.~Bernier}
\altaffiliation{Present address: Department of Physics, University of the Western Cape, P/B X17, Bellvill, ZA-7535, South Africa}
\affiliation{TRIUMF, 4004 Wesbrook Mall, Vancouver, British Columbia V6T 2A3, Canada}
\affiliation{Department of Physics and Astronomy, University of British Columbia, Vancouver, British Columbia V6T 1Z4, Canada}
\author{H.~Bidaman}
\affiliation{Department of Physics, University of Guelph, Guelph, Ontario N1G 2W1, Canada}
\author{V.~Bildstein}
\affiliation{Department of Physics, University of Guelph, Guelph, Ontario N1G 2W1, Canada}
\author{M.~Bowry}
\altaffiliation{Present address: School of Computing, Engineering and Physical Sciences, University of the West of Scotland, Paisley PA1 2BE, United Kingdom}
\affiliation{TRIUMF, 4004 Wesbrook Mall, Vancouver, British Columbia V6T 2A3, Canada}
\author{D.S.~Cross}
\affiliation{Department of Chemistry, Simon Fraser University, Burnaby, British Columbia V5A 1S6, Canada}
\author{M.R.~Dunlop}
\affiliation{Department of Physics, University of Guelph, Guelph, Ontario N1G 2W1, Canada}
\author{R.~Dunlop}
\affiliation{Department of Physics, University of Guelph, Guelph, Ontario N1G 2W1, Canada}
\author{A.B.~Garnsworthy}
\affiliation{TRIUMF, 4004 Wesbrook Mall, Vancouver, British Columbia V6T 2A3, Canada}
\author{P.E.~Garrett}
\affiliation{Department of Physics, University of Guelph, Guelph, Ontario N1G 2W1, Canada}
\author{J.~Henderson}
\altaffiliation{Present address: Lawrence Livermore National Laboratory, Livermore, California 94550, USA}
\affiliation{TRIUMF, 4004 Wesbrook Mall, Vancouver, British Columbia V6T 2A3, Canada}
\author{J.~Measures}
\affiliation{TRIUMF, 4004 Wesbrook Mall, Vancouver, British Columbia V6T 2A3, Canada}
\affiliation{University of Surrey, Guildford GU2 7XH, United Kingdom}
\author{B.~Olaizola}
\affiliation{TRIUMF, 4004 Wesbrook Mall, Vancouver, British Columbia V6T 2A3, Canada}
\author{K.~Ortner}
\affiliation{Department of Chemistry, Simon Fraser University, Burnaby, British Columbia V5A 1S6, Canada}
\author{J.~Park}
\altaffiliation{Present address: Department of Physics, Lund University, 22100 Lund, Sweden}
\affiliation{TRIUMF, 4004 Wesbrook Mall, Vancouver, British Columbia V6T 2A3, Canada}
\affiliation{Department of Physics and Astronomy, University of British Columbia, Vancouver, British Columbia V6T 1Z4, Canada}
\author{C.M.~Petrache}
\affiliation{Centre de Sciences Nucl\'{e}aire et Sciences de la Mati\`{e}re, CNRS/IN2P3, Universit\'{e} Paris-Saclay, Orsay, France}
\author{J.L.~Pore}
\altaffiliation{Present address: Lawrence Berkeley National Laboratory, Berkeley, California 94720, USA}
\affiliation{Department of Chemistry, Simon Fraser University, Burnaby, British Columbia V5A 1S6, Canada}
\author{K.~Raymond}
\affiliation{Department of Chemistry, Simon Fraser University, Burnaby, British Columbia V5A 1S6, Canada}
\author{J.K.~Smith}
\altaffiliation{Present Address: Department of Physics, Pierce College, Puyallup, Washington 98374, USA}
\affiliation{TRIUMF, 4004 Wesbrook Mall, Vancouver, British Columbia V6T 2A3, Canada}
\author{D.~Southall}
\altaffiliation{Present address: Department of Physics, University of Chicago, Chicago, IL 60637, USA}
\affiliation{TRIUMF, 4004 Wesbrook Mall, Vancouver, British Columbia V6T 2A3, Canada}
\author{C.E.~Svensson}
\affiliation{Department of Physics, University of Guelph, Guelph, Ontario N1G 2W1, Canada}
\author{M.~Ticu}
\affiliation{Department of Chemistry, Simon Fraser University, Burnaby, British Columbia V5A 1S6, Canada}
\author{J.~Turko}
\affiliation{Department of Physics, University of Guelph, Guelph, Ontario N1G 2W1, Canada}
\author{K.~Whitmore}
\affiliation{Department of Chemistry, Simon Fraser University, Burnaby, British Columbia V5A 1S6, Canada}
\author{T.~Zidar}
\affiliation{Department of Physics, University of Guelph, Guelph, Ontario N1G 2W1, Canada}

\date{\today}

\begin{abstract}
The $\beta$-decay of neutron-rich $^{129}$In into $^{129}$Sn was studied using the GRIFFIN spectrometer at the ISAC facility at TRIUMF. The study observed the half-lives of the ground state and each of the $\beta$-decaying isomers. The level scheme of $^{129}$Sn has been expanded with thirty-one new $\gamma$-ray transitions and nine new excited levels, leading to a re-evaluation of the $\beta$-branching ratios and level spin assignments. The observation of the $\beta$-decay of the (29/2$^{+}$) 1911-keV isomeric state in $^{129}$In is reported for the first time, with a branching ratio of 2.0(5)$\%$.

\end{abstract}

\pacs{Valid PACS appear here}
\maketitle

%
\thispagestyle{empty}

\section{Introduction\label{sec:intro}}
The region around the doubly magic $^{132}_{\texttt{ }50}$Sn$_{82}$ nucleus is replete with critical information required for nuclear structure models and astrophysical applications \cite{Cottle, Bazin}. 
This isotope region is a key input to the nuclear shell model and the theoretical frameworks required to establish a working predictive and descriptive model of nuclei, and as a result it has been the focus of a series of theoretical studies \cite{Andreozzi, Otsuka2001, Otsuka2005}.

In nuclear astrophysics there is also a need for information on this region due to the importance of the $A=130$ elemental abundance peak \cite{Beun, Surman}. The rapid neutron-capture process (\textit{r}-process) is responsible for the generation of isotopes heavier than iron in stellar environments \cite{Burbidge, Cameron1958}, and is shown to have key waiting points at the magic shell closures in the vicinity of the tin isotopes \cite{Cameron1983, Mumpower}.

The $^{129}$Sn nucleus, three neutrons removed from the $N=82$ shell closure, is important for studying the effects of single neutron excitations and other shell effects, as evidenced by the sixty years of study it has undergone. Several production mechanisms have been used in order to study $^{129}$In, and its $\beta^{-}$ daughter, $^{129}$Sn, including fission \cite{Pinston}, $\beta^{-}$-decay \cite{Aleklett, DeGeer, Spanier, Huck, Gausemel}, $\beta$n-decay \cite{Warner, Rudstam} and internal transition decay \cite{Genevey, Lozeva}. Though the information on the transitions and energy levels in this daughter nucleus is plentiful, there are virtually no definitive spin or parity assignments for the levels above the 3/2$^{+}$ ground state, the 11/2$^{-}$ 35-keV isomer and the (1/2)$^{+}$ 315-keV excited state. 

In order to study and increase the available information on this key nucleus,  $^{129}$Sn, high-efficiency $\gamma$-ray spectroscopy and coincidence techniques were used to uncover new transitions, new decay patterns and new levels, providing more input information for state-of-the-art theoretical models. 
\section{Experiment\label{sec:exp}}
The Isotope Separator and ACcelerator (ISAC) facility of TRIUMF \cite{ISAC} employs the Isotope Separation On-Line (ISOL) technique in order to produce radioactive isotope beams \cite{ISOL}. Isotopes are generated by bombarding a uranium carbide (UC$_{{x}}$) target with a 9.8~$\mu$A beam of 480~MeV protons, provided by the main 520-MeV cyclotron \cite{Cyclotron}. The relevant isotopes are selectively ionized for extraction using the Ion-Guide Laser Ion Source (IGLIS) \cite{IGLIS}, in order to reduce any isobaric contamination. The ionized species are then passed through the high-resolution mass spectrometer ($M/\delta M \sim2000$) \cite{MassSep}, in order to produce an isotopically clean beam. Once extracted, the desired $^{129}$In radioactive isotope beam is transported to the experimental station. The $\beta^{-}$ decay of the $^{129}$In isotope to $^{129}$Sn was observed using the Gamma-Ray Infrastructure For Fundamental Investigations of Nuclei (GRIFFIN) \cite{GRIFFINSpec, FullGRIFFIN, GRIFFIN2019}\\
\indent The GRIFFIN array is a state-of-the-art, high-resolution $\gamma$-ray spectrometer, equipped with sixteen high-purity germanium (HPGe) clover detectors for the identification of $\gamma$-rays \cite{Rizwan}. Each of the sixteen HPGe clover detectors contains four crystals, making a total of 64 crystals that can detect $\gamma$-rays, allowing for analyses to be carried out in single crystal or addback modes \cite{FullGRIFFIN, Rizwan}. For this experiment, the SCintillating Electron-Positron Tagging ARray (SCEPTAR) \cite{FullGRIFFIN} was placed at the centre of GRIFFIN, in order to provide tagging for $\beta$ particles. A cycling mylar tape station, the focus of which is at the centre of SCEPTAR, provides a continuous implantation spot and aids in the removal of contaminants. An implantation cycle can be set to optimize observation of the decay of interest. During the course of this experiment, a mix of the $\beta^{-}$ decaying isomers of $^{129}$In were implanted at a rate of $\sim$5000 particles per second, with a beam composition of approximately 41\% in the 9/2$^{+}$ ground state, $^{129}$In$^{gs}$, 54\% in the (1/2$^{-}$) 459-keV $^{129}$In$^{m1}$ isomer, 3\% in the (23/2$^{-}$) 1630-keV $^{129}$In$^{m2}$ isomer and 1\% in the (29/2$^{+}$) 1911-keV $^{129}$In$^{m3}$ isomer, neglecting the uncertainty in the ground state branch of the (1/2$^{-}$) isomer. \\
\indent The GRIFFIN array was arranged in its high-efficiency configuration, where the HPGe clover detectors were positioned 11 cm away from the implantation spot \cite{FullGRIFFIN}. A 20 mm Delrin shield was put in place around SCEPTAR to minimize Bremsstrahlung radiation from high energy $\beta^{-}$ particles. The experimental campaign for $^{129}$In consisted of running the tape system in consecutive 21.5-second cycles, with 1.5 s for tape move, 5 s for background collection, 10 s for isotope implantation and 5 s for isotope decay. The total run duration was 2.75 hrs, for a total of 460 cycles with 6.29 $\times 10^{7}$ addback singles events, and 1.81 $\times 10^{7}$ coincidence events collected during the run time.
The combination of GRIFFIN and SCEPTAR allowed for the correlated observation of $\gamma$-rays in coincidence with the emitted $\beta$ particles, in a 500~ns coincidence window, in order to tag on the specific $^{129}$In decays. Furthermore, $\gamma$-$\gamma$ coincidences, with a 500~ns coincidence window, were used for the verification of transitions and decay patterns through the excited levels of the $^{129}$Sn daughter.\\
\indent The energy and efficiency calibrations for GRIFFIN were done using a series of standard sources of $^{56}$Co, $^{60}$Co, $^{133}$Ba, $^{152}$Eu, allowing for a calibration to be made in the range between 81 keV and 3.6 MeV. Coincidence summing corrections were done by constructing a $\gamma$-$\gamma$ matrix with detectors positioned at 180$^{\circ}$ of each other to correct for real coincidence summing, a methodology established for GRIFFIN in Ref. \cite{GRIFFIN2019}. Transitions were also verified to be real transitions rather than sum or escape peaks. 

\section{Results\label{sec:results}}
\subsection{Transitions and Levels in $^{129}$Sn\label{sub:tinLevelScheme}}

\begin{figure}[h!]
\includegraphics[width=\columnwidth]{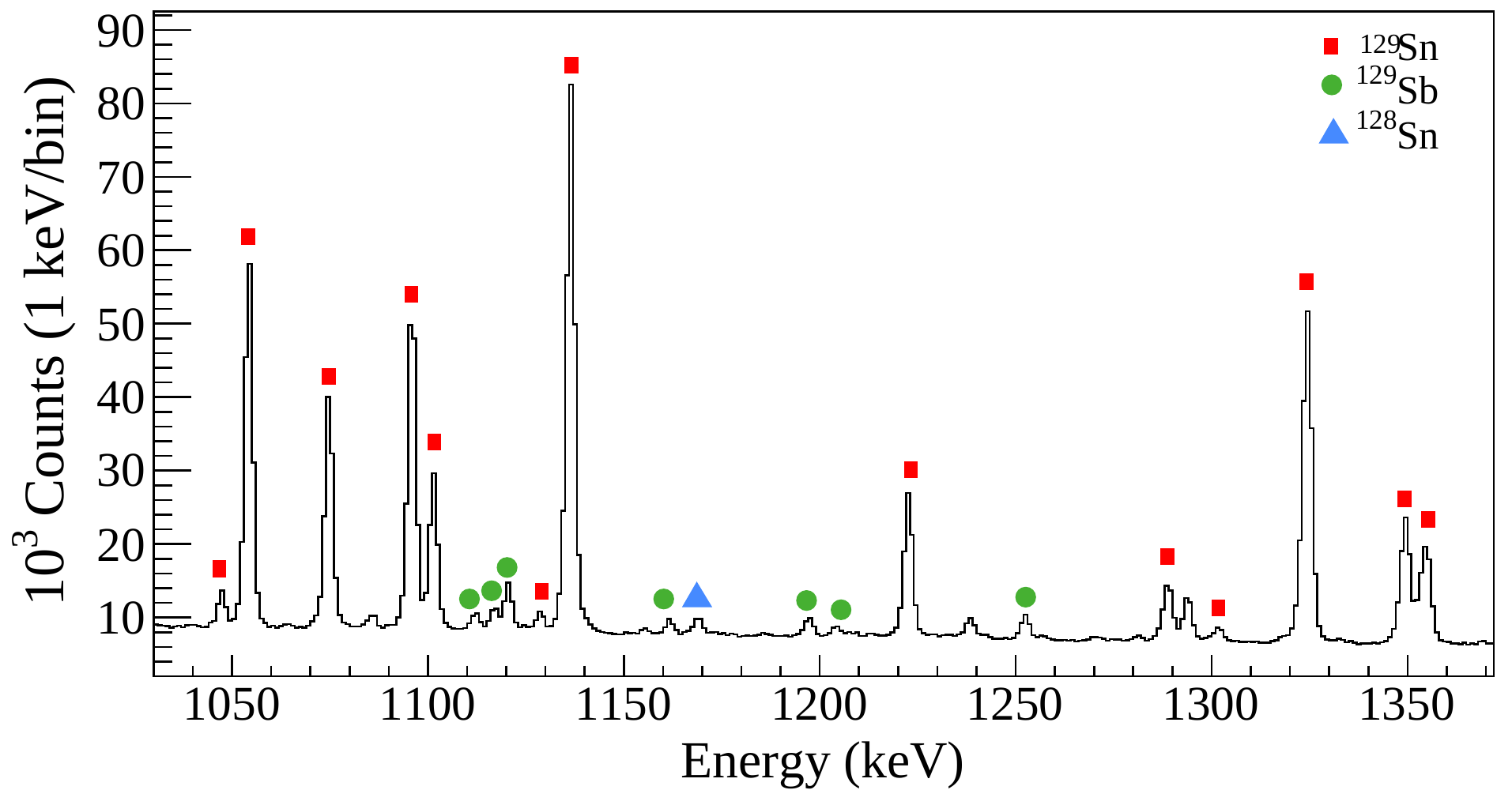}
\caption{$\beta$-gated $\gamma$-ray spectrum, in addback mode, showing various known transitions in $^{129}$Sn (red squares). Transitions in $^{129}$Sb ($\beta$ decay daughter of $^{129}$Sn; green circles) and $^{128}$Sn ($\beta$n daugther of $^{129}$In; blue triangles) are also observed. \label{fig:aabSpectrum}}
\end{figure}

\begin{figure}[h]
\includegraphics[width=\columnwidth]{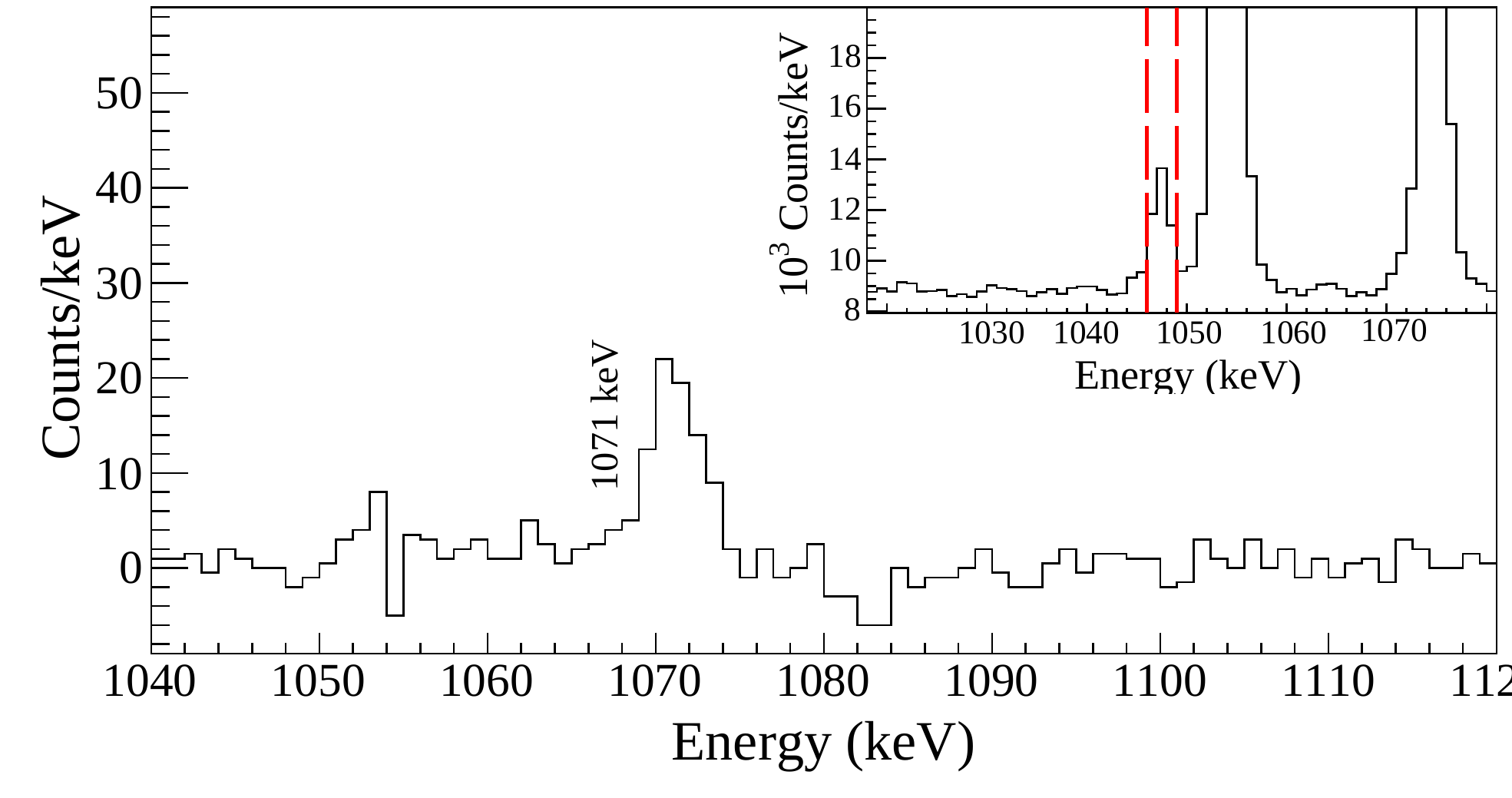}
\caption{A $\gamma$-ray spectrum, in addback mode, showing evidence for the 1071-keV $\gamma$-ray, depopulating the 2118-keV state, in coincidence with the 1047-keV $\gamma$-ray, depopulating the 1047-keV state of $^{129}$Sn. The 1071-keV transition lies on the shoulder of a much more intense transition at 1075-keV, necessitating gating from below to obtain its relative intensity. The inset shows the gate on 1047~keV used to produce the spectrum. \label{fig:Inset1071}}
\end{figure}

\begin{figure}[h]
\includegraphics[width=\columnwidth]{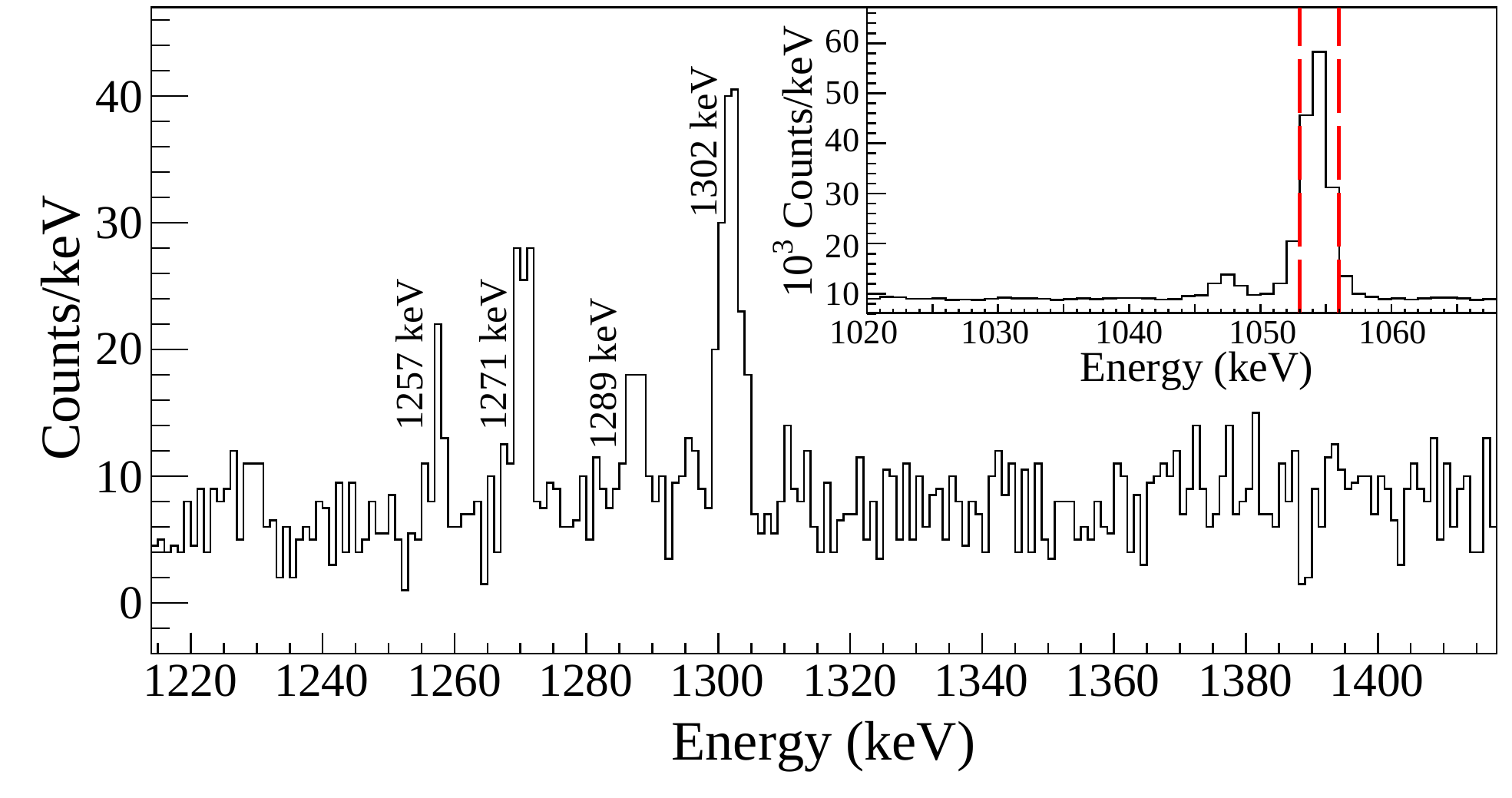}
\caption{A $\gamma$-ray spectrum, in addback mode, showing evidence for the 1271-keV $\gamma$-ray, depopulating the new state at 2326-keV, in coincidence with the 1054-keV $\gamma$-ray, depopulating the 1054-keV state in $^{129}$Sn. Though weak, this transition is visible in the ungated $\gamma$-ray spectrum, but it is much more clear in the 1054-keV coincidence spectrum. This coincidence also confirms its placement in the level scheme. The transition at 1257 keV is also newly observed, while those at 1289 and 1302 keV are known in $^{129}$Sn. The inset shows the gate on 1054~keV used to produce the spectrum. \label{fig:Inset1054}}
\end{figure}

The $Q_{\beta}$ value for the $^{129}$In $\beta$-decay to $^{129}$Sn is 7.769(19) MeV and the neutron separation energy, $S_{n}$, for $^{129}$Sn is 5.316(26) MeV \cite{NNDC-129In}. Gamma-ray transitions were investigated up to the neutron separation energy. From the analysis of this data set, all but two of the transitions currently reported for the $^{129}$Sn nucleus were observed \cite{NNDC-129In}. There were also 31 newly observed transitions and 9 newly observed excited states in the $^{129}$Sn nucleus, never observed through the $\beta^{-}$ decay of its $^{129}$In parent or otherwise. Figure \ref{fig:aabSpectrum} shows a  portion of the $\beta$-gated $\gamma$-ray spectrum observed in this work; transitions in the $^{129}$Sn of interest, along with transitions in the $\beta$ granddaughter, $^{129}$Sb, and in the $^{129}$In $\beta$n daughter, $^{128}$Sn are identified. Figures~\ref{fig:Inset1071} and \ref{fig:Inset1054} demonstrate the mechanism used to establish new transitions. Figure~\ref{fig:Inset1071} shows the gating from below method used to determine several intensities, in this case that of the 1071-keV transition. The intensity of this transition required gating from below on the 1047-keV transition, while Figure~\ref{fig:Inset1054} shows the $\gamma$-ray spectrum resulting from a gate placed on the 1054-keV transition, where coincident transitions at 1257, 1271 and 1302 keV were observed.

Table \ref{tab:gammaTable} summarizes the energy levels observed in this work, along with the transitions from each level, the final state, the relative intensity with respect to the highest intensity 2118-keV $\gamma$-ray, and the $\gamma$-ray branching ratio. The table also compares the branching ratios for each of the known $\gamma$-rays appearing in the evaluation by Timar, Elekes and Singh \cite{NNDC-129In}.  
With the exception of seven $\gamma$-rays at 146, 278, 280, 1071, 1096, 1586, and 2371 keV, all transition intensities were obtained from the addback singles spectrum. The seven $\gamma$-rays mentioned required directly gating from below in order to fit their energy and intensity values, as demonstrated in Figure~\ref{fig:Inset1071}. The other new transitions were observed in the $\gamma$-ray spectrum and their placement confirmed through coincidence gating, as in shown in Figure~\ref{fig:Inset1054}.

Though most of the branching ratios observed in the course of this work are in good agreement with the work of Gausemel \textit{et al.} \cite{Gausemel}, there are some notable discrepancies. These are attributed to major differences in the conditions of the two experiments. The work by Gausemel \textit{et al.} utilized three germanium detectors, while the present work made use of all 16 HPGe clover detectors available to the GRIFFIN array, allowing for more efficient coincidence detection. Furthermore, some of the transitions were observed with branching ratios down to $10^{-4}$, pushing the limits of the detection mechanisms available to the previous experiment.
The $\beta$-decay of the (29/2$^{+}$) 1911-keV isomer in $^{129}$In to the (27/2$^{-}$) 2552-keV state in $^{129}$Sn was observed for the first time. This 2552-keV state was previously observed in the fission study done by Lozeva \textit{et al.} \cite{Lozeva}.
The transitions from the isomeric states at 1762 keV and 1803 keV to lower-lying states in the level scheme, 19.7 keV and 41.0 keV, respectively, were not observed. However transitions feeding into these states were present in the data, confirming the placement of the levels, within uncertainty.

\newpage
\onecolumngrid
\thispagestyle{empty}
\noindent
\begin{ThreePartTable}
  \begin{TableNotes}
    \item[*] new spin assignment
  \end{TableNotes}
\LTcapwidth=\textwidth
    \begin{longtable*}{@{\extracolsep{\fill}}c|cccc|c|c||c}
         \caption[table]{Energy levels and transitions observed in $^{129}$Sn, following the $\beta^{-}$ decay of $^{129}$In. All intensities are normalized to the most intense transition at 2118 keV, from the (7/2$^{+}$) 2118-keV state to the 3/2$^{+}$ ground state. The values calculated in this experiment are compared to those present in the Evaluated Nuclear Structure Data File search (ENSDF) database of the National Nuclear Data Center (NNDC). The level spins and parities are adopted from Ref. \cite{NNDC-129In}, unless otherwise stated.\label{tab:gammaTable}}\\
      \hline \hline 
      \multicolumn{7}{c||}{This work}&{ENSDF} \\ 
     \cline{1-8} $E$ level (keV)&$E_{\gamma}$ (keV)&$J^{\pi}_{i}$&$J^{\pi}_{f}$&$E_{f}$ (keV)&Relative $I_{\gamma}$&$BR_{\gamma}$&$BR_{\gamma}$
    \\ \hline \hline 
    \endfirsthead
\caption[]{\em{(Continued).}} \\
      \hline \hline	
     \multicolumn{7}{c||}{This work}&{ENSDF} \\ 
     \cline{1-8} $E$ level (keV)&$E_{\gamma}$ (keV)&$J^{\pi}_{i}$&$J^{\pi}_{f}$&$E_{f}$ (keV)&Relative $I_{\gamma}$&$BR_{\gamma}$&$BR_{\gamma}$
    \\ \hline \hline
\endhead
\hline
\endfoot
\endlastfoot
0	&		&	3/2$^{+}$	&		&		&		&		&		\\	\hline
35.2(2)	&		&	11/2$^{-}$	&		&		&		&		&		\\	\hline
315.1(2)	&	315.4(2)	&	(1/2$^{+}$)	&	3/2$^{+}$	&	0	&	0.611(5)	&	100	&	100	\\	\hline
763.7(1)	&	728.5(2)	&	(9/2$^{-}$)	&	11/2$^{-}$	&	35.2(2)	&	0.163(2)	&	100	&	100	\\	\hline
769.1(1)	&	769.3(2)	&	(5/2$^{+}$)	&	3/2$^{+}$	&	0	&	0.297(3)	&	100	&	100	\\	\hline
1043.9(1)	&	280.4(2)$^{\dagger}$	&	(7/2$^{-}$)	&	(9/2$^{-}$)	&	763.7(1)	&	0.0068(8)	&	3.6(5)	&	4.3(6)	\\	
	&	1008.5(2)	&	(7/2$^{-}$)	&	11/2$^{-}$	&	35.2(2)	&	0.186(2)	&	100(1)	&	100	\\	\hline
1047.0(2)	&	278.0(2)$^{\dagger}$	&	(7/2$^{+}$)	&	(5/2$^{+}$)	&	769.1(1)	&	0.0080(10)	&	95(7)	&	68(16)	\\	
	&	1047.4(2)	&	(7/2$^{+}$)	&	3/2$^{+}$	&	0	&	0.0086(4)	&	100(5)	&	100(8)	\\	\hline
1054.3(2)	&	285.2(2)	&	(7/2$^{+}$)	&	(5/2$^{+}$)	&	769.1(1)	&	0.036(2)	&	34(1)	&	33(2)	\\	
	&	1054.4(2)	&	(7/2$^{+}$)	&	3/2$^{+}$	&	0	&	0.1050(10)	&	100.0(5)	&	100(7)	\\	\hline
1171.5(3)	&	1136.4(2)	&	(15/2$^{-}$)	&	11/2$^{-}$	&	35.2(2)	&	0.1698(14)	&	100	&	100(7)	\\	\hline
1222.4(2)	&	175.5(4)	&	(3/2$^{+}$)	&	(7/2$^{+}$)	&	1047.0(2)	&	0.0025(6)	&	5(1)	&	2.2(13)	\\	
	&	907.3(2)	&	(3/2$^{+}$)	&	(1/2$^{+}$)	&	315.1(2)	&	0.0394(7)	&	86(1)	&	74(5)	\\	
	&	1222.6(2)	&	(3/2$^{+}$)	&	3/2$^{+}$	&	0	&	0.0461(5)	&	100.0(9)	&	100(7)	\\	\hline
1288.6(2)	&	519.5(7)	&	(3/2$^{+}$)	&	(5/2$^{+}$)	&	769.1(1)	&	0.007(3)	&	32(14)	&	36(6)$^{\mathsection}$	\\	
	&	973.6(2)	&	(3/2$^{+}$)	&	(1/2$^{+}$)	&	315.1(2)	&	0.0193(9)	&	88(4)	&	80(5)$^{\mathsection}$	\\	
	&	1288.8(2)	&	(3/2$^{+}$)	&	3/2$^{+}$	&	0	&	0.0220(4)	&	100(2)	&	100(6)$^{\mathsection}$	\\	\hline
1359.5(3)	&	1324.4(2)	&	(13/2$^{-}$)	&	11/2$^{-}$	&	35.2(2)	&	0.1306(13)	&	100	&	100	\\	\hline
1455.2(2)	&	1455.0(2)	&	(5/2$^{+}$)	&	3/2$^{+}$	&	0	&	0.0306(9)	&	100	&	100	\\	\hline
1534.4(2)	&	480.2(2)	&	(7/2$^{-}$, 9/2$^{+}$)	&	(7/2$^{+}$)	&	1054.3(2)	&	0.0137(11)	&	100(8)	&	100(9)	\\	
	&	765.0(3)	&	(7/2$^{-}$, 9/2$^{+}$)	&	(5/2$^{+}$)	&	769.1(1)	&	0.0037(11)	&	27(8)	&	69(6)	\\	
	&	1499.1(2)	&	(7/2$^{-}$, 9/2$^{+}$)	&	11/2$^{-}$	&	35.2(2)	&	0.0114(7)	&	84(5)	&	99(7)	\\	\hline
1607.3(3)	&	553.1(3)	&	(7/2 – 11/2)$^{*}$	&	(7/2$^{+}$)	&	1054.3(2)	&	0.0037(6)	&	58(9)	&		\\	
	&	843.4(3)	&	(7/2 – 11/2)$^{*}$	&	(7/2$^{+}$)	&	763.7(1)	&	0.0064(5)	&	100(7)	&		\\	\hline
1613.6(3)	&	1613.4(2)	&	(7/2$^{+}$)$^{\ddagger}$	&	3/2$^{+}$	&	0	&	0.0359(5)	&	100	&	100(6)	\\	\hline
1688.3(3)	&	919.0(3)	&	(7/2$^{-}$, 9/2$^{+}$)$^{*}$	&	(5/2$^{+}$)	&	769.1(1)	&	0.0027(2)	&	100(9)	&		\\	
	&	1653.0(3)	&	(7/2$^{-}$, 9/2$^{+}$)$^{*}$	&	11/2$^{-}$	&	35.2(2)	&	0.0025(3)	&	92(12)	&		\\	\hline
1701.0(2)	&	657.3(2)	&	(7/2$^{-}$)	&	(7/2$^{-}$)	&	1043.9(1)	&	0.0043(3)	&	45(4)	&	75(6)$^{\mathsection}$	\\	
	&	932.0(2)	&	(7/2$^{-}$)	&	(5/2$^{+}$)	&	769.1(1)	&	0.0094(4)	&	100(4)	&	100(9)$^{\mathsection}$	\\	
	&	937.4(2)	&	(7/2$^{-}$)	&	(9/2$^{-}$)	&	763.7(1)	&	0.0075(5)	&	79(6)	&	95(9)$^{\mathsection}$	\\	
	&	1665.6(3)	&	(7/2$^{-}$)	&	11/2$^{-}$	&	35.2(2)	&	0.0027(7)	&	28(8)	&		\\	\hline
1741.9(3)	&	382.4(2)	&	(15/2$^{+}$)	&	(13/2$^{-}$)	&	1359.5(3)	&	0.1235(12)	&	77.2(6)	&	75(4)	\\	
	&	570.4(2)	&	(15/2$^{+}$)	&	(15/2$^{-}$)	&	1171.5(3)	&	0.160(2)	&	100.0(7)	&	100(6)	\\	\hline
1853.3(2)	&	318.0(6)	&	(7/2, 9/2)	&	(7/2$^{-}$, 9/2$^{+}$)	&	1534.4(2)	&	0.0073(3)	&	48(2)	&	32(4)	\\	
	&	799.4(2)	&	(7/2, 9/2)	&	(7/2$^{+}$)	&	1054.3(2)	&	0.0153(9)	&	100(6)	&	100(8)	\\	
	&	806.3(4)	&	(7/2, 9/2)	&	(7/2$^{+}$)	&	1047.0(2)	&	0.0016(4)	&	11(3)	&		\\	
	&	1085.7(6)	&	(7/2, 9/2)	&	(5/2$^{+}$)	&	769.1(1)	&	0.0046(3)	&	30(2)	&		\\	\hline
1865.1(1)	&	330.9(3)	&	(7/2$^{+}$)	&	(7/2$^{-}$, 9/2$^{+}$)	&	1534.4(2)	&	0.0060(14)	&	0.7(2)	&	0.66(5)	\\	
	&	411.2(6)	&	(7/2$^{+}$)	&	(5/2$^{+}$)	&	1455.2(2)	&	0.0083(5)	&	1.03(6)	&	0.34(4)	\\	
	&	576.1(3)	&	(7/2$^{+}$)	&	(3/2$^{+}$)	&	1288.6(2)	&	0.0009(2)	&	0.11(3)	&	0.39(3)	\\	
	&	821.4(2)	&	(7/2$^{+}$)	&	(7/2$^{-}$)	&	1043.9(1)	&	0.0173(5)	&	2.14(5)	&	2.22(1)	\\	
	&	1095.9(2)$^{\dagger}$	&	(7/2$^{+}$)	&	(5/2$^{+}$)	&	769.1(1)	&	0.081(5)	&	9.9(6)	&	8.5(11)	\\	
	&	1101.4(2)	&	(7/2$^{+}$)	&	(9/2$^{-}$)	&	763.7(1)	&	0.0435(7)	&	5.36(7)	&	5.7(4)	\\	
	&	1830.6(3)	&	(7/2$^{+}$)	&	11/2$^{-}$	&	35.2(2)	&	0.0039(5)	&	0.48(6)	&	0.32(7)	\\	
	&	1864.8(2)	&	(7/2$^{+}$)	&	3/2$^{+}$	&	0	&	0.812(7)	&	100.0(6)	&	100(7)	\\	\hline
1906.2(2)	&	1906.2(2)	&	(7/2)	&	3/2$^{+}$	&	0	&	0.0093(4)	&	100	&	100	\\	\hline
2023.6(4)	&	969.2(3)	&	(7/2 – 11/2)$^{*}$	&	(7/2$^{+}$)	&	1054.3(2)	&	0.0123(7)	&	100	&		\\	\hline
2118.3(1)	&	212.2(3)	&	(7/2$^{+}$)	&	(7/2)	&	1906.2(2)	&	0.0045(6)	&	0.45(6)	&	0.64(5)	\\	
	&	253.1(3)	&	(7/2$^{+}$)	&	(7/2$^{+}$)	&	1865.1(1)	&	0.0057(4)	&	0.57(4)	&	0.08(2)	\\	
	&	265.5(3)	&	(7/2$^{+}$)	&	(7/2, 9/2)	&	1853.3(2)	&	0.0036(4)	&	0.36(4)	&	0.35(5)	\\	
	&	583.6(2)	&	(7/2$^{+}$)	&	(7/2$^{-}$, 9/2$^{+}$)	&	1534.4(2)	&	0.0144(7)	&	1.44(7)	&		\\	
	&	662.9(2)	&	(7/2$^{+}$)	&	(5/2$^{+}$)	&	1455.2(2)	&	0.0125(3)	&	1.25(3)	&	1.22(8)	\\	
	&	829.9(2)	&	(7/2$^{+}$)	&	(3/2$^{+}$)	&	1288.6(2)	&	0.0048(3)	&	0.48(3)	&	0.6(10)	\\	
	&	1071.0(2)$^{\dagger}$	&	(7/2$^{+}$)	&	(7/2$^{+}$)	&	1047.0(2)	&	0.0006(1)	&	0.06(1)	&	0.2(10)	\\	
	&	1074.7(2)	&	(7/2$^{+}$)	&	(7/2$^{-}$)	&	1043.9(1)	&	0.0666(7)	&	6.66(7)	&	6.1(4)	\\	
	&	1349.5(2)	&	(7/2$^{+}$)	&	(5/2$^{+}$)	&	769.1(1)	&	0.0511(8)	&	5.11(8)	&	4.6(3)	\\	
	&	1354.7(2)	&	(7/2$^{+}$)	&	(9/2$^{-}$)	&	763.7(1)	&	0.0417(11)	&	4.2(1)	&	2.9(3)	\\	
	&	2083.0(3)	&	(7/2$^{+}$)	&	11/2$^{-}$	&	35.2(2)	&	0.0033(4)	&	0.33(4)	&	0.42(4)	\\	
	&	2118.3(2)	&	(7/2$^{+}$)	&	3/2$^{+}$	&	0	&	1	&	100.0(6)	&	100(7)	\\	\hline
2277(1)	&	474.0(2)	&	(21/2)	&	(23/2)$^{+}$	&	1803(1)	&	0.0220(5)	&	100(2)	&	100(8)	\\	
	&	514.8(3)	&	(21/2)	&	(19/2)$^{+}$	&	1762(1)	&	0.0081(6)	&	37(6)	&	30(3)	\\	\hline
2326.1(4)	&	1270.5(6)	&	(7/2, 9/2$^{+}$)$^{*}$	&	(7/2$^{+}$)	&	1054.3(2)	&	0.0012(3)	&	82(21)	&		\\	
	&	1558.1(4)	&	(7/2, 9/2$^{+}$)$^{*}$	&	(5/2$^{+}$)	&	769.1(1)	&	0.0015(3)	&	100(18)	&		\\	\hline
2406(1)	&	604.4(5)	&	(23/2$^{-}$)	&	(23/2$^{+}$)	&	1803(1)	&	0.0090(4)	&	100	&	100	\\	\hline
2552(1)	&	145.5(3)$^{\dagger}$	&	(27/2$^{-}$)	&	(23/2$^{-}$)	&	2406(1)	&	0.0011(2)	&	100	&	100	\\	\hline
2568.0(3)	&	2252.9(3)	&	(1/2, 3/2)$^{*}$	&	(1/2)$^{+}$	&	315.1(2)	&	0.0009(3)	&	93(30)	&		\\	
	&	2568.0(3)	&	(1/2, 3/2)$^{*}$	&	3/2$^{+}$	&	0	&	0.0010(6)	&	100(56)	&		\\	\hline
2606.2(2)	&	1150.9(3)	&	(1/2, 3/2)$^{*}$	&	(5/2$^{+}$)	&	1455.2(2)	&	0.0002(2)	&	6(4)	&		\\	
	&	1384.2(3)	&	(1/2, 3/2)$^{*}$	&	(3/2$^{+}$)	&	1222.4(2)	&	0.0035(4)	&	90(11)	&		\\	
	&	2290.5(3)	&	(1/2, 3/2)$^{*}$	&	(1/2)$^{+}$	&	315.1(2)	&	0.0013(2)	&	33(6)	&		\\	
	&	2606.9(4)	&	(1/2, 3/2)$^{*}$	&	3/2$^{+}$	&	0	&	0.0039(4)	&	100(9)	&		\\	\hline
2791.0(3)	&	1736.6(3)	&	(7/2, 9/2$^{+}$)	&	(7/2$^{+}$)	&	1054.3(2)	&	0.0018(3)	&	22(3)	&	36(24)	\\	
	&	2021.9(2)	&	(7/2, 9/2$^{+}$)	&	(5/2$^{+}$)	&	769.1(1)	&	0.0082(3)	&	100(3)	&	100(7)	\\	\hline
2836.0(2)	&	718.0(3)	&	(7/2$^{+}$, 9/2$^{+}$)	&	(7/2$^{+}$)	&	2118.3(1)	&	0.0026(5)	&	5(1)	&		\\	
	&	1301.8(2)	&	(7/2$^{+}$, 9/2$^{+}$)	&	(7/2$^{-}$, 9/2$^{+}$)	&	1534.4(2)	&	0.0058(8)	&	12(2)	&	10.2(8)	\\	
	&	1781.4(2)	&	(7/2$^{+}$, 9/2$^{+}$)	&	(7/2$^{+}$)	&	1054.3(2)	&	0.0490(6)	&	100(1)	&	100(7)	\\	
	&	1791.4(3)	&	(7/2$^{+}$, 9/2$^{+}$)	&	(7/2$^{-}$)	&	1043.9(1)	&	0.0031(4)	&	6.3(7)	&	7.2(7)	\\	
	&	2066.5(2)	&	(7/2$^{+}$, 9/2$^{+}$)	&	(5/2$^{+}$)	&	769.1(1)	&	0.0299(6)	&	61(1)	&	57(4)	\\	
	&	2072.9(3)	&	(7/2$^{+}$, 9/2$^{+}$)	&	(9/2$^{-}$)	&	763.7(1)	&	0.0003(1)	&	0.6(2)	&	3.5(9)	\\	\hline
2981.9(2)	&	863.8(4)	&	(7/2$^{+}$)	&	(7/2$^{+}$)	&	2118.3(1)	&	0.0007(3)	&	3(1)	&		\\	
	&	1128.7(2)	&	(7/2$^{+}$)	&	(7/2, 9/2)	&	1853.3(2)	&	0.0036(8)	&	18(4)	&		\\	
	&	1927.6(3)	&	(7/2$^{+}$)	&	(7/2$^{-}$)	&	1054.3(2)	&	0.0015(3)	&	8(1)	&		\\	
	&	2212.6(2)	&	(7/2$^{+}$)	&	(5/2$^{+}$)	&	769.1(1)	&	0.0201(5)	&	100(2)	&	100(6)	\\	
	&	2980.7(7)	&	(7/2$^{+}$)	&	3/2$^{+}$	&	0	&	0.0009(2)	&	4.5(8)	&	9.2(18)	\\	\hline
3079.3(3)	&	2035.6(3)	&	(3/2$^{-}$)	&	(7/2$^{-}$) 	&	1043.9(1)	&	0.0045(7)	&	79(11)	&	90(9)	\\	
	&	2764.0(2)	&	(3/2$^{-}$)	&	(1/2)$^{+}$	&	315.1(2)	&	0.0057(3)	&	100(5)	&	100(8)	\\	\hline
3140.3(2)	&	1526.1(3)	&	(7/2$^{+}$)	&	(7/2$^{+}$)	&	1613.6(3)	&	0.0020(5)	&	17(4)	&		\\	
	&	2094.0(3)	&	(7/2$^{+}$)	&	(7/2$^{+}$)	&	1047.0(2)	&	0.0041(4)	&	34(3)	&		\\	
	&	2371.1(3)$^{\dagger}$	&	(7/2$^{+}$)	&	(5/2$^{+}$)	&	769.1(1)	&	0.0045(3)	&	37(3)	&	16.2(18)	\\	
	&	2376.4(3)	&	(7/2$^{+}$)	&	(9/2$^{-}$)	&	763.7(1)	&	0.0025(6)	&	20(5)	&	30(3)	\\	
	&	3140.1(2)	&	(7/2$^{+}$)	&	3/2$^{+}$	&	0	&	0.0123(3)	&	100(2)	&	100(8)	\\	\hline
3393.9(4)	&	3078.7(3)	&	(1/2, 3/2)	&	(1/2)$^{+}$	&	315.1(2)	&	0.0077(3)	&	100	&	100	\\	\hline
3446.7(4)	&	2683.0(3)	&	(7/2 – 11/2)$^{*}$	&	(9/2$^{-}$)	&	763.7(1)	&	0.0005(2)	&	100	&		\\	\hline
3581.8(3)	&	1257.0(6)	&	(7/2, 9/2$^{+}$)$^{*}$	&	(7/2, 9/2)	&	2326.1(4)	&	0.0013(2)	&	90(17)	&		\\	
	&	2527.1(3)	&	(7/2, 9/2$^{+}$)$^{*}$	&	(7/2$^{+}$)	&	1054.3(2)	&	0.0013(3)	&	93(23)	&		\\	
	&	2812.7(8)	&	(7/2, 9/2$^{+}$)$^{*}$	&	(5/2$^{+}$)	&	769.1(1)	&	0.0014(4)	&	100(25)	&		\\	
	&	2818.4(5)	&	(7/2, 9/2$^{+}$)$^{*}$	&	(9/2$^{-}$)	&	763.7(1)	&	0.0010(4)	&	69(27)	&		\\	\hline
3590.4(1)	&	1889.5(2)	&	(3/2$^{-}$)	&	(7/2$^{-}$) 	&	1701.0(2)	&	0.0106(5)	&	23(1)	&	23(2)	\\	
	&	1977.0(2)	&	(3/2$^{-}$)	&	(7/2$^{+}$)	&	1613.6(3)	&	0.0113(5)	&	24(1)	&	21(2)	\\	
	&	2301.7(2)	&	(3/2$^{-}$)	&	(3/2$^{+}$) 	&	1288.6(2)	&	0.0178(3)	&	38.0(5)	&	30(2)	\\	
	&	2367.9(2)	&	(3/2$^{-}$)	&	(3/2$^{+}$) 	&	1222.4(2)	&	0.0292(4)	&	62.3(7)	&	51(4)	\\	
	&	2546.2(2)	&	(3/2$^{-}$)	&	(7/2$^{-}$) 	&	1043.9(1)	&	0.0469(5)	&	100(1)	&	100(7)	\\	
	&	3276.0(2)	&	(3/2$^{-}$)	&	(1/2)$^{+}$ 	&	315.1(2)	&	0.0440(6)	&	94(1)	&	63(4)	\\	
	&	3589.7(3)	&	(3/2$^{-}$)	&	3/2$^{+}$	&	0	&	0.0058(5)	&	12(1)	&	12.6(11)	\\	\hline
3993(1)	&	1586.3(3)$^{\dagger}$	&	(21/2$^{-}$)	&	(23/2$^{-}$)	&	2406(1)	&	0.011(6)	&	7.2(1)	&		\\	
	&	1715.9(2)	&	(21/2$^{-}$)	&	(21/2)	&	2277(1)	&	0.0286(7)	&	17.6(4)	&	16.4(14)	\\	
	&	2189.8(2)	&	(21/2$^{-}$)	&	(23/2$^{+}$)	&	1803(1)	&	0.162(2)	&	100.0(7)	&	100(7)	\\	
	&	2230.8(2)	&	(21/2$^{-}$)	&	(19/2$^{+}$)	&	1762(1)	&	0.0577(9)	&	35.5(5)	&	41(2)	\\	\hline
4136.6(3)	&	2847.8(2)	&	(1/2, 3/2)$^{*}$	&	(3/2$^{+}$)	&	1288.6(2)	&	0.0026(1)	&	100(3)	&		\\	
	&	2915.5(5)	&	(1/2, 3/2)$^{*}$	&	(3/2$^{+}$)	&	1222.4(2)	&	0.0019(3)	&	73(12)	&		\\			
\hline \hline
	\multicolumn{8}{l}{$^{*}$ Spin assignment for new levels, based on $\beta$ and $\gamma$ decay systematics.} \\
	\multicolumn{8}{l}{$^{\dagger}$ Intensity calculated from coincidences.} \\
	\multicolumn{8}{l}{$^{\ddagger}$ Revised spin assignment for known levels, based on $\beta$ and $\gamma$ decay systematics.} \\
	\multicolumn{8}{l}{$^{\mathsection}$ Based on ENSDF information: Weighted average between $^{129}$In$^{gs}$ and $^{129}$In$^{m1}$.} \\
    \end{longtable*}
	\end{ThreePartTable}
\twocolumngrid
\noindent
\subsection{Decay of the  9/2$^{+}$ $^{129}\textrm{In}$ ground state \label{sub:groundState}}
\subsubsection{Half-life of $^{129}$In$^{gs}$ \label{subsub:halfLife-gs}}
\begin{figure}[h!]
\includegraphics[width=\columnwidth]{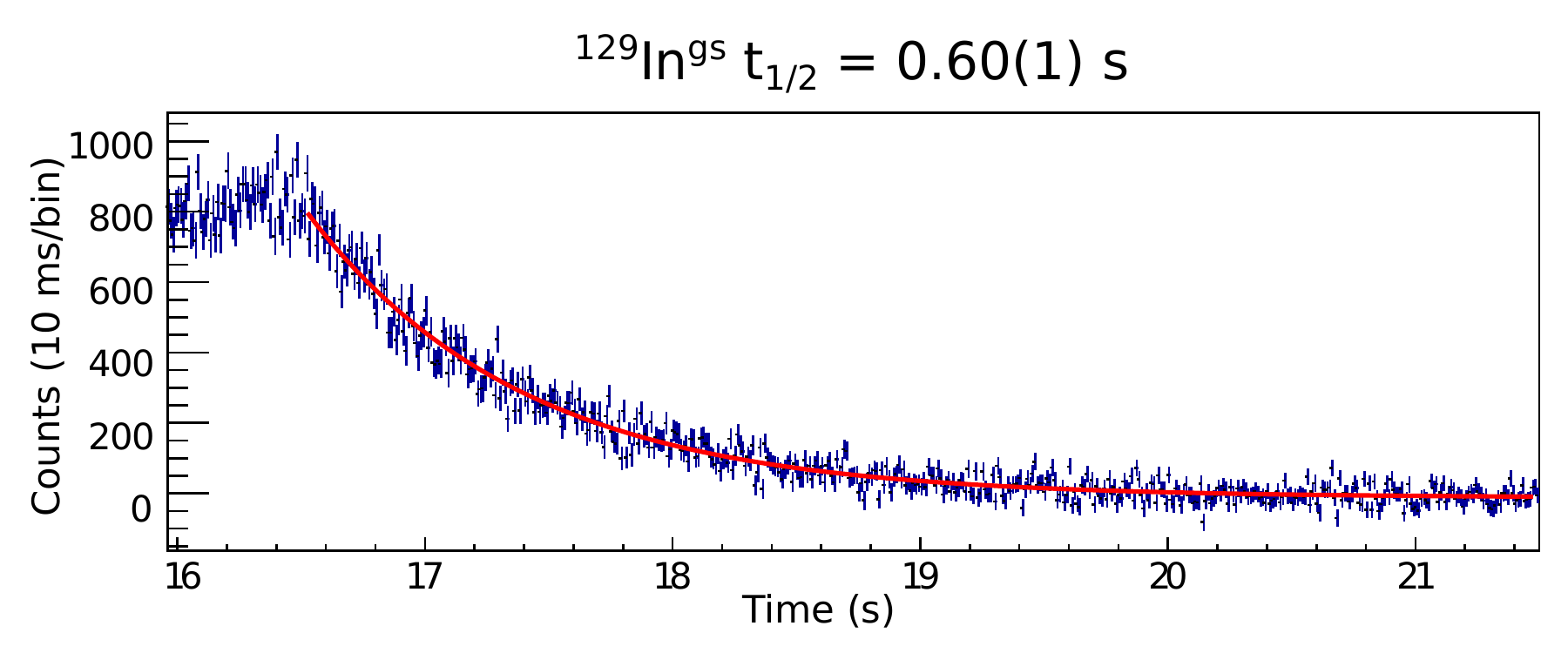}
\caption{A spectrum of total counts as a function of cycle time, representing 39 transitions associated with the $^{129}$In ground state decay into states in $^{129}$Sn. The fit, seen in red, returned a value of $t_{1/2} = 0.60(1)$ s. The reduced $\chi^{2}$ for this fit is 1.3.\label{fig:halfLife-gs}}
\end{figure}

Plotting the intensity of a $\gamma$-ray as a function of cycle time allows for the measurement of isotope half-life; a spectrum can be generated and fit with a characteristic decay formula, from which the half-life can then be extracted. 
To improve statistics, background-corrected timing gates on 39 $\gamma$-rays --- shown in Table \ref{tab:haflLifePeaks-gs} --- associated with the decay of the $^{129}$In ground state into $^{129}$Sn were summed and the counts as a function of cycle time fit using a standard exponential decay, as seen in Figure \ref{fig:halfLife-gs}. The fit returned a half-life of $t_{1/2} = 0.60(1)$ s, in agreement with $t_{1/2} = 0.611(5)$ s quoted in the evaluation by Timar, Elekes and Singh \cite{NNDC-129In}. A chop analysis, which involved a change in the width of the timing window for the fit, was conducted to check for any rate dependent effects on the half-life; no discernible effects were observed. The weighted average value between the evaluted half-life of 0.611(5) s and the observed half-live of 0.60(1) s is calculated to be 0.609(4) s, where the uncertainty has been increase by the $\sqrt{\chi^{2}}$.

\begin{table}[h!]
\centering
 \caption{Transitions used to build the half-life plot shown in Figure \ref{fig:halfLife-gs}. These were identified as transitions from states populated by the ground state of $^{129}$In. \label{tab:haflLifePeaks-gs}}
 \begin{tabular}{r|r|r|r}
    \hline \hline 
 \multicolumn{4}{c}{Transitions (keV)}\\ \hline
212.2(3)	&	662.9(2)	&	1349.5(2)	&	2021.9(2)	\\
253.1(3)	&	765.0(3)	&	1354.7(2)	&	2066.5(2)	\\
265.5(3)	&	799.4(2)	&	1499.1(2)	&	2072.9(3)	\\
278.0(2)	&	821.4(2)	&	1613.4(2)	&	2083.0(3)	\\
285.2(2)	&	829.9(2)	&	1736.6(3)	&	2118.3(2)	\\
318.0(6)	&	1054.4(2)	&	1781.4(2)	&	2212.6(2)	\\
330.9(3)	&	1074.7(2)	&	1791.4(3)	&	2376.4(3)	\\
411.2(6)	&	1095.9(2)	&	1830.6(3)	&	2980.7(7)	\\
480.2(2)	&	1101.4(2)	&	1864.8(2)	&	3140.1(2)	\\
576.1(3)	&	1301.8(2)	&	1906.2(2)	&		\\
     \hline \hline
\end{tabular}
\end{table}

\subsubsection{ $\beta$-feeding and logft values \label{sub:gs-feeding}}

\begin{figure*}[ht!]
\centering
\centering
\includegraphics[width=\textwidth]{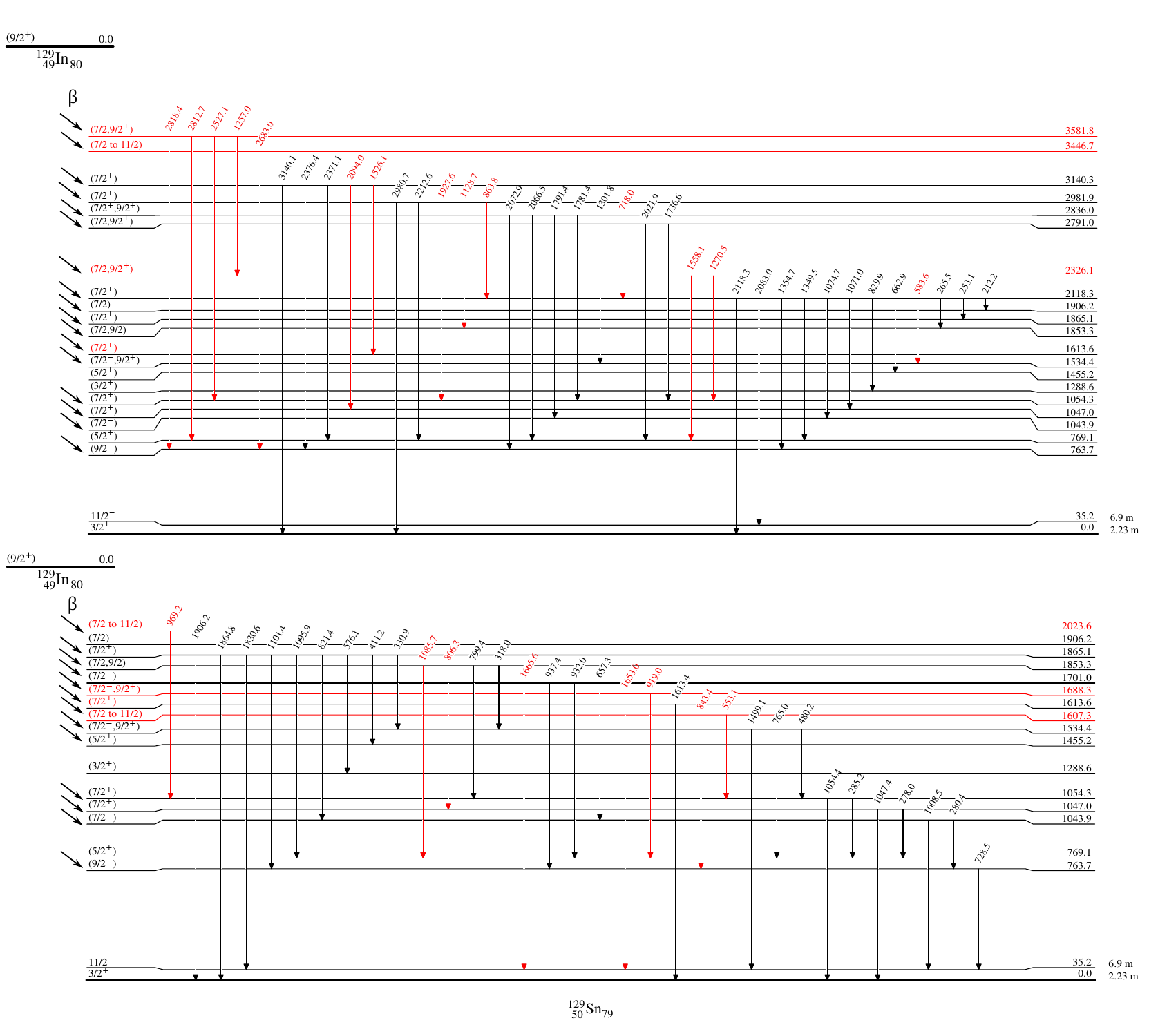}
\caption{The level scheme of $^{129}$Sn, populated through the $\beta$-decay of the ground state of $^{129}$In, showing the high-lying states (top) and the low-lying states (bottom). The colour (red) represents new transitions and levels found in this work. For the case of the 1614-keV state, the coloured (7/2$^{+}$) spin indicates a new spin assignment to a previously observed level. The half-lives of the ground state and the $^{129}$Sn$^{m1}$ 35-keV isomer are 2.23(4) min and 6.9(1) min, respectively, as given by Timar, Elekes and Singh \cite{NNDC-129In}. Information about $\gamma$-ray intensity and their uncertainties can be found in Table \ref{tab:gammaTable}. \label{fig:lvlScheme-gs}}
\end{figure*} 

Several new transitions and new levels from the $\beta$-decay of the ground state of $^{129}$In into excited states of $^{129}$Sn were observed. Figure \ref{fig:lvlScheme-gs} shows the $\gamma$-rays observed in $^{129}$Sn due to this decay. Newly observed transitions and levels are coloured (red), along with their proposed spin assignments. 

Table \ref{tab:betaFeeding-gs} lists the states that are populated by the ground state along with the $\beta$-feeding intensities and the log\textit{ft} values calculated in this work, together with a comparison to the work of Gausemel \textit{et al.} \cite{Gausemel}. The electron conversion coefficients for low-energy $\gamma$-rays are taken into account when doing these calculations, using the BrIcc utility available through the NNDC \cite{Bricc}. The data observed by Gausemel \textit{et al.} showed direct $\beta$-feeding to the 35-keV isomeric state on the order of $<10\%$. The $\beta$-feeding values obtained in the present work are normalized to reflect 100$\%$ feeding to excited states.

Spin assignments for the newly observed levels are proposed based on the $\beta$-decay selection rules and $\gamma$-ray systematics. The newly observed states at 1607~keV, 2024~keV and 3447~keV are tentatively assigned spins between (7/2) and (11/2), as they are observed to decay to states with proposed spins between 7/2 and 9/2, while decays to states with spins of 3/2 or lower were not observed. 

The new state at 1688 keV decays to the 11/2$^{-}$ 35-keV and (5/2$^{+}$) 769-keV levels such that the spin and parity of this state can be restricted to (7/2$^{-}$, 9/2$^{+}$). The new states at 2326 keV and 3582 keV are observed to decay to states with tentative spins between (5/2$^{+}$) and 9/2$^{-}$, such that their spins can be restricted to (7/2, 9/2$^{+}$). The log\textit{ft} values calculated between the (9/2$^{+}$) $^{129}$In$^{gs}$ and the above mentioned states, shown in Table \ref{tab:betaFeeding-gs}, are all consistent with the allowed or first-forbidden decays implied by the proposed spin assignments.
\begin{table}[ht!]
\centering
 \caption{The $\beta$-feeding intensities and log\textit{ft} values for states in $^{129}$Sn, observed through the $\beta$-decay of the (9/2$^{+}$) $^{129}$In$^{gs}$ state and calculated with the weighted average half-life of 0.609(4) s. Columns denoted by Ref. \cite{Gausemel} contain values established in the work of Gausemel \textit{et al.}. The $\beta$-feeding values have been normalized to reflect 100\% feeding of excited states.\label{tab:betaFeeding-gs}}
 \begin{tabular}{c|cc|cc}
    \hline \hline
\multirow{2}{*}{$E_{x}$ (keV)} & \multicolumn{2}{c}{$I_{\beta}$ (\%)} & \multicolumn{2}{c}{log\textit{ft}} \\ 
 \cline{2-5}  & This work & Ref. \cite{Gausemel} & This work & Ref. \cite{Gausemel} \\ \hline
763.7(1)	&	2.03(8)	&	2.1(4)	&	6.43(2)	&	6.4(1)	\\
1043.9(1)	&	1.95(8)	&	2.0(4)	&	6.36(2)	&	6.4(1)	\\
1047.0(2)	&	0.30(4)	&	0.35(6)	&	7.17(6)	&	7.1(1)	\\
1054.3(2)	&	1.62(10)	&	2.1(3)	&	6.44(3)	&	6.33(7)	\\
1534.4(2)	&	0.60(9)	&	0.46(6)	&	6.73(7)	&	6.85(6)	\\
1607.3(3)	&	0.14(2)	&		&	7.33(7)	&		\\
1613.6(3)	&	0.88(3)	&		&	6.54(2)	&		\\
1688.2(3)	&	0.20(2)	&		&	7.15(4)	&		\\
1701.0(2)	&	0.52(5)	&	0.24(2)	&	6.74(5)	&	7.08(4)	\\
1853.3(2)	&	0.85(5)	&	0.76(6)	&	6.47(3)	&	6.53(4)	\\
1865.1(1)	&	37.6(3)	&	36(2)	&	4.83(1)	&	4.85(3)	\\
1906.2(2)	&	0.18(3)	&	0.13(3)	&	7.13(8)	&	7.3(1)	\\
2023.6(4)	&	0.48(3)	&		&	6.67(3)	&		\\
2118.3(1)	&	46.9(3)	&	49(3)	&	4.64(1)	&	4.63(3)	\\
2326.1(4)	&	0.06(2)	&		&	7.49(14)	&		\\
2791.0(3)	&	0.39(15)	&	0.47(9)	&	6.48(2)	&	6.4(1)	\\
2836.0(2)	&	3.53(5)	&	3.36(15)	&	5.51(1)	&	5.54(2)	\\
2981.9(2)	&	1.01(4)	&	0.74(5)	&	5.99(2)	&	6.14(3)	\\
3140.3(2)	&	0.99(4)	&	0.67(4)	&	5.94(2)	&	6.11(3)	\\
3446.7(4)	&	0.020(9)	&		&	7.5(2)	&		\\
3581.8(3)	&	0.19(3)	&		&	6.46(7)	&		\\
     \hline \hline
\end{tabular}
\end{table}

Previous work using $\gamma$-ray information \cite{Gausemel, NNDC-129In} assigned a spin of  between (1/2) and (7/2$^{+}$) for the 1614-keV level, with no detectable $\beta$-feeding from the (1/2$^{-}$) $^{129}$In$^{m1}$ parent. Gausemel \textit{et al.} observed a transition at 1977~keV, from the (3/2$^{-}$) 3590-keV state to the 1614-keV state, which was observed in the present work and is shown in Figure \ref{fig:lvlScheme-m1}. A new transition, at 1526-keV, was also observed from the (7/2$^{+}$) state at 3140 keV, casting doubt on the possibility of a 1/2 spin. 
Furthermore, the 1614-keV level is observed to have a direct $\beta$-feeding component, equivalent to double the intensity of the 1977-keV transition from the 3590-keV state observed both by Gausemel \textit{et al.} and in this work, indicating that the 1614-keV level is most likely fed by the (9/2$^{+}$) $^{129}$In$^{gs}$ and has a spin of (7/2$^{+}$). This spin assignment is corroborated by the 6.54(2) log\textit{ft} value shown in Table \ref{tab:betaFeeding-gs}, which is consistent with an allowed transition. 
\vspace{-10pt}
\subsection{Decay of (1/2$^{-}$) $^{129}\textrm{In}^{m1}$ \label{sub:m1State}}
\subsubsection{Half-life of $^{129}$In$^{m1}$\label{subsub:halfLife-m1}}
\begin{figure}[h!]
\includegraphics[width=\columnwidth]{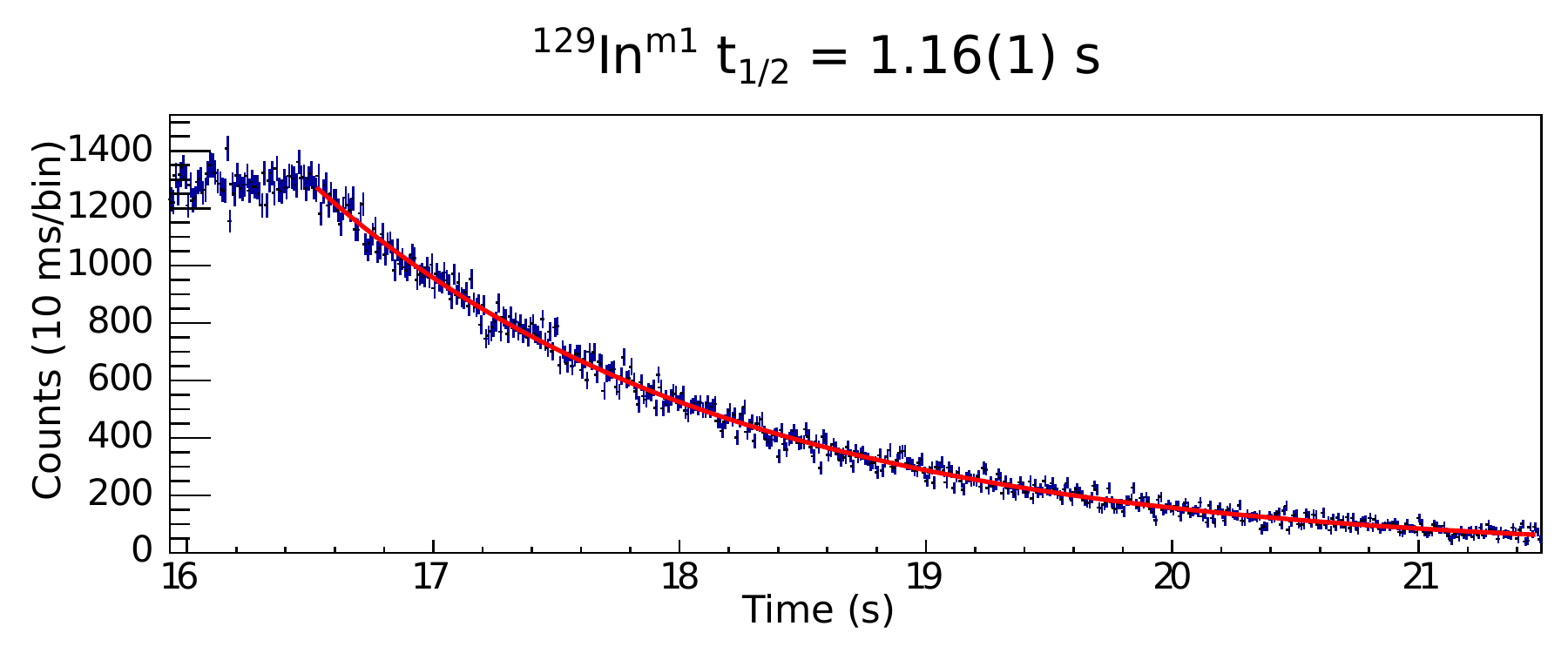}
\caption{A spectrum of total counts as a function of cycle time, representing twelve transitions associated with the decay of $^{129}$In$^{m1}$ into states in $^{129}$Sn. The fit, seen in red, returned a value of $t_{1/2} = 1.16(1)$ s. The reduced $\chi^{2}$ for this fit is 1.2.\label{fig:halfLife-m1}}
\end{figure}
The twelve $\gamma$-ray transitions used to generate the half-life spectrum shown in Figure \ref{fig:halfLife-m1} are listed in Table \ref{tab:haflLifePeaks-m1}. The fit to the data returned a half-life of $t_{1/2} = 1.16(1)$ s. A chop analysis was carried out and no systematic effects were observed. The present result is a factor of three more precise than the value 1.23(3) s quoted in the evaluation by Timar, Elekes and Singh \cite{NNDC-129In}. The weighted average of these two values is 1.17(2) s, with its uncertainty increased by $\sqrt{\chi^{2}}$.
\begin{figure*}[ht!]
\begin{minipage}[c]{\textwidth}
\includegraphics[width=\textwidth]{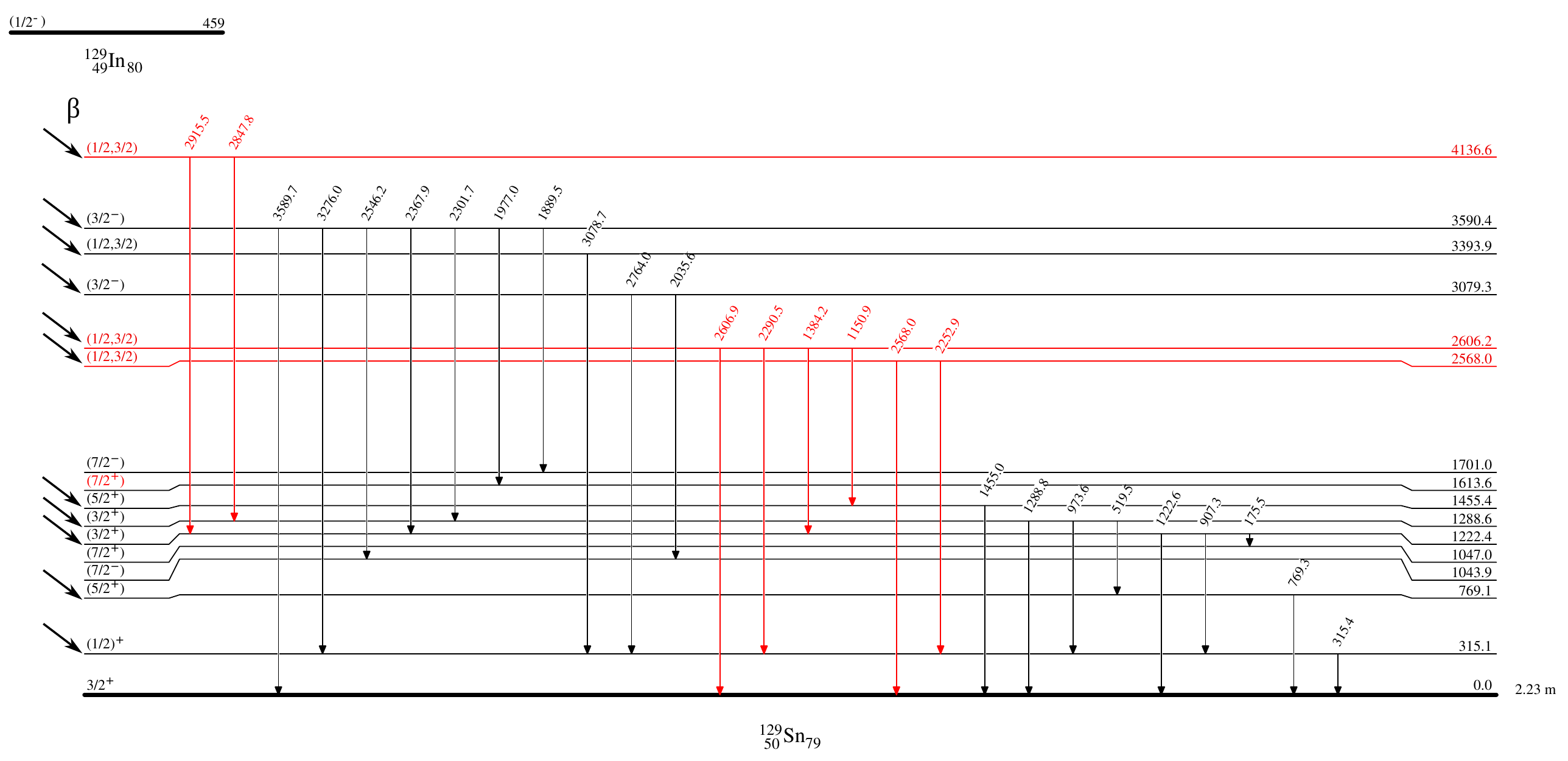}
\caption{The level scheme of $^{129}$Sn, populated through the $\beta$-decay of the (1/2$^{-}$) 459-keV isomer of $^{129}$In. The colour (red) represents new transitions and levels found in this work. The half-life of the $^{129}$Sn ground state is 2.23(4) min, as given by Timar, Elekes and Singh \cite{NNDC-129In}. Information about $\gamma$-ray intensity and their uncertainties can be found in Table \ref{tab:gammaTable}.\label{fig:lvlScheme-m1}}
\end{minipage}
\end{figure*} 

\begin{table}[h!]
\centering
 \caption{Transitions used to build the half-life plot shown in Figure \ref{fig:halfLife-m1}. These were identified as transitions from states populated by the $^{129}In^{m1}$ isomer. \label{tab:haflLifePeaks-m1}}
 \begin{tabular}{r|r|r}
    \hline \hline 
 \multicolumn{3}{c}{Transitions (keV)}\\ \hline
175.5(4)	&	1889.5(2)	&	2764.0(2)	\\
315.4(2)	&	2035.6(3)	&	3078.7(3)	\\
907.3(2)	&	2367.9(2)	&	3276.0(2)	\\
1222.6(2)	&	2546.2(2)	&	3589.7(3)	\\
     \hline \hline
\end{tabular}
\end{table}

\subsubsection{$\beta$-feeding and logft values \label{subsub:m1-feeding}}

The $^{129}$In$^{m1}$ isomer was observed to decay to known states in $^{129}$Sn, as well as to three newly observed states. Figure \ref{fig:lvlScheme-m1} shows the $\gamma$-rays observed in this decay; the spins of the three new levels populated in $^{129}$Sn were determined from $\beta$-feeding and $\gamma$-ray systematics. Table \ref{tab:betaFeeding-m1} summarizes the $\beta$-feeding intensities and the log\textit{ft} values obtained in the present work, using the weighted half-life value of 1.17(2) s, compared with the results of Gausemel \textit{et al.} \cite{Gausemel} who also observed a 77(15)$\%$ direct $\beta$-feeding to the 3/2$^{+}$ ground state in $^{129}$Sn; the present $\beta$-feeding intensities have been scaled to represent the remaining 23$\%$ of observed $\beta$-feeding to excited states. 

The new states at 2568 keV and 2606 keV are observed to decay to the (1/2)$^{+}$ 315-keV state; therefore they are likely to have a spin of either 1/2 or 3/2, and be fed by the (1/2$^{-}$) $^{129}$In$^{m1}$. The 4137-keV state decays to the 1289-keV and 1222-keV states, both of which have tentative spin assignments of (3/2$^{+}$), indicating that this state is likely to have a 1/2 or 3/2 spin. The log\textit{ft} values for these states, shown in Table \ref{tab:betaFeeding-m1}, are consistent with either the allowed or first-forbidden transitions expected from the (1/2$^{-}$) $^{129}$In$^{m1}$ to the respective states in $^{129}$Sn.
\begin{table}[h!]
\centering
 \caption{The $\beta$-feeding intensities and log\textit{ft} values, calculated, for states in $^{129}$Sn, observed through the $\beta$-decay of the (1/2$^{-}$) $^{129}$In$^{m1}$ isomer and calculated with the weighted average half-life of 1.17(2) s. The values calculated in this work are compared to the values calculated by Gausemel \textit{et al.} \cite{Gausemel}. \label{tab:betaFeeding-m1}}
 \begin{tabular}{c|cc|cc}
    \hline \hline
\multirow{2}{*}{$E_{x}$ (keV)} & \multicolumn{2}{c}{$I_{\beta}$ (\%)} & \multicolumn{2}{c}{log\textit{ft}} \\ 
 \cline{2-5}  & This work & Ref. \cite{Gausemel} & This work & Ref. \cite{Gausemel}\\ \hline
315.1(2)		&	14.3(2)	&	15.1(13)	&	6.10(1)	&	6.10(4)	\\
769.1(1)	$^{\dagger}$	&	0.9(2)	&		&	9.31(9)	&		\\
1222.4(2)		&	1.52(4)	&	1.56(14)	&	6.83(2)	&	6.85(4)	\\
1288.6(2)		&	0.63(9)	&	0.58(7)	&	7.20(7)	&	7.26(6)	\\
1455.2(2)	$^{\dagger}$	&	0.28(3)	&		&	9.55(5)	&		\\
2568.0(3)		&	0.06(2)	&		&	7.86(15)	&		\\
2606.2(2)		&	0.25(2)	&		&	7.19(4)	&		\\
3079.3(3)		&	0.29(2)	&	0.42(3)	&	6.96(4)	&	6.82(4)	\\
3393.9(4)		&	0.219(8)	&	0.22(2)	&	6.96(2)	&	6.98(5)	\\
3590.4(1)		&	4.70(6)	&	5.10(17)	&	5.55(1)	&	5.54(2)	\\
4136.6(3)		&	0.127(9)	&		&	6.88(4)	&		\\
     \hline \hline
          	\multicolumn{5}{l}{$^{\dagger}$ Unique 1st forbidden}\\
\end{tabular}
\end{table}

The excess feeding into the (5/2$^{+}$) states at 769 keV and 1455 keV has been attributed to direct $\beta$-feeding from the (1/2$^{-}$) $^{129}$In$^{m1}$ to these states, rather than to unobserved transitions feeding these levels from higher levels populated by either the $\beta$-decay of the (9/2$^{+}$) $^{129}$In$^{gs}$ or the (1/2$^{-}$) $^{129}$In$^{m1}$.
The log\textit{ft} values for the 769-keV and 1455-keV states, calculated with feeding from $^{129}$In$^{m1}$ are 9.31(9) and 9.55(5), respectively, consistent with unique first-forbidden transitions, supporting the spin assignments for these states, given as (5/2$^{+}$).

\subsection{Decay of (23/2$^{-}$) $^{129}\textrm{In}^{m2}$ \label{sub:m2State}}

\subsubsection{Half-life of $^{129}$In$^{m2}$ \label{subsub:halfLife-m2}}
\begin{figure}[h!]
\includegraphics[width=\columnwidth]{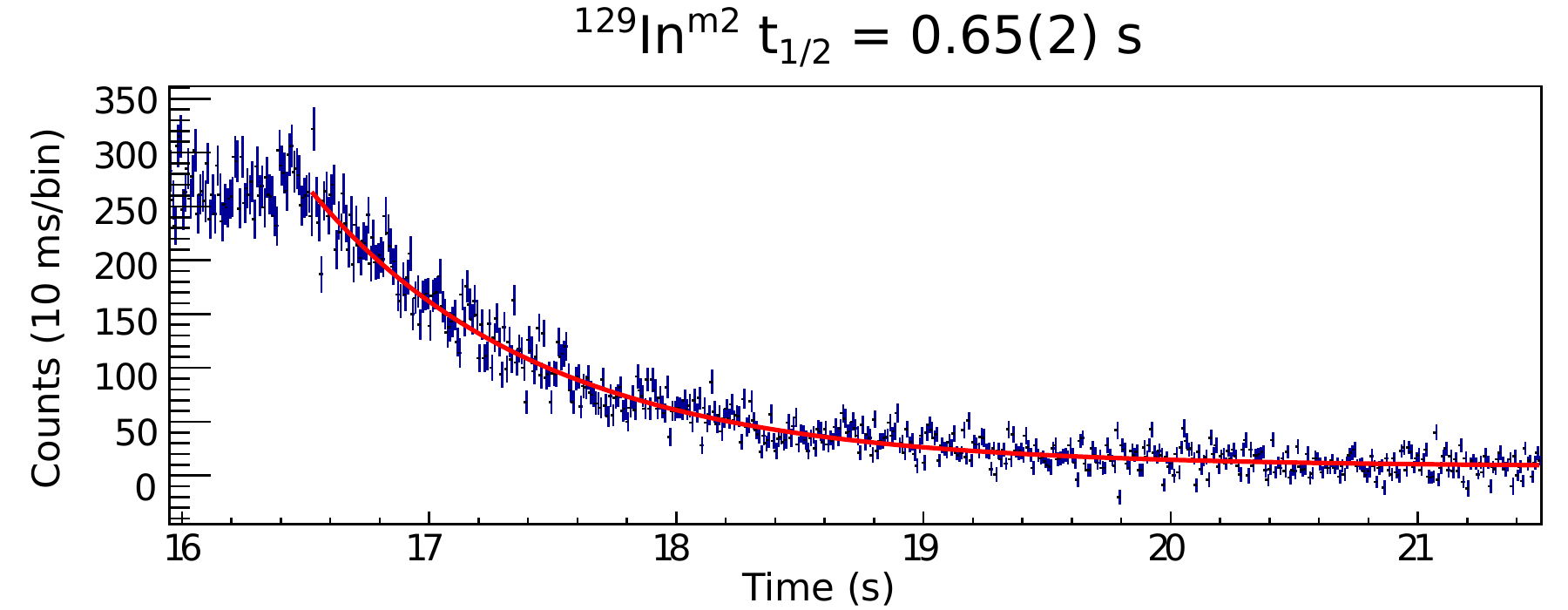}
\caption{A spectrum of total counts as a function of cycle time, representing four transitions associated with $^{129}$In$^{m2}$ decay into states in $^{129}$Sn. The fit, seen in red, returned a value of $t_{1/2} = 0.65(1)$ s.  The reduced $\chi^{2}$ for this fit is 2.2.\label{fig:halfLife-m2}}
\end{figure}

Only four viable $\gamma$-ray transitions, at 382, 515, 2190 and 2231~keV, could be used in the half-life fitting of the $^{129}$In$^{m2}$ isomer. The spectrum and the fit are show in Figure \ref{fig:halfLife-m2}. The adopted value for the half-life of the (23/2$^{-}$) $^{129}$In$^{m2}$ is $t_{1/2} = 0.65(2)$ s. This result, which includes a systematic uncertainty associated with the chop analysis, has an error of a factor of five times smaller than the established value, quoted by Timar, Elekes and Singh \cite{NNDC-129In}, as $t_{1/2} = 0.67(10)$ s. The weighted average between the half-life in the evalution and the half-life observed in this work is 0.65(2) s, with the uncertainty increased by $\sqrt{\chi^{2}}$. 

\subsubsection{$\beta$-feeding and logft values \label{subsub:m2-feeding}}
\begin{figure*}[ht!]
\begin{minipage}[c]{\textwidth}
\includegraphics[width=0.8\columnwidth]{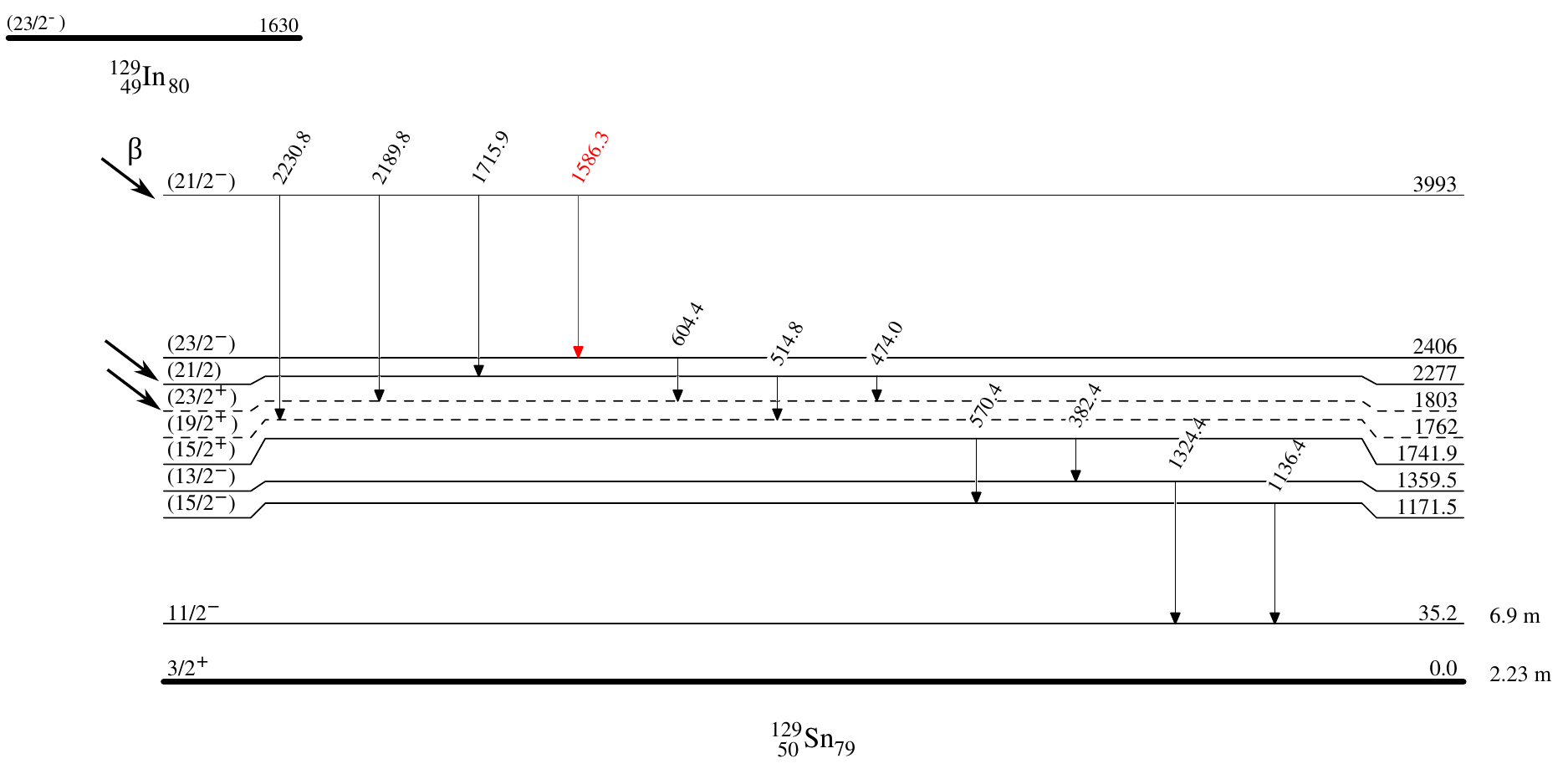}
\caption{The level scheme of $^{129}$Sn, populated through the $\beta$-decay of the (23/2$^{-}$) $^{129}$In. The colour (red) represents new transitions and levels found in this work. The half-life of the $^{129}$Sn ground state is 2.23(4) min, as given by Timar, Elekes and Singh \cite{NNDC-129In}. Information about $\gamma$-ray intensity and their uncertainties can be found in Table \ref{tab:gammaTable}. The dashed lines presented two known states, at 1762 keV and 1803 keV, whose energies can only be inferred in this work from to feeding from above.\label{fig:lvlScheme-m2}}
\end{minipage}  
\end{figure*}

The level scheme associated with the $\beta$-decay of the $^{129}$In$^{m2}$ is shown in Figure \ref{fig:lvlScheme-m2}. Table \ref{tab:betaFeeding-m2} shows the $\beta$ feeding intensities and log\textit{ft} values obtained in this work, along with a comparison to those observed by Gausemel et al.\cite{Gausemel}. One new transition was observed in the present work. The branching ratios measured for the decay of the 3993-keV (21/2$^{-}$) level to the 1762-, 1803- and 2277-keV levels and for the decay of the 2277-keV level to the 1762- and 1803-keV levels are in reasonable agreement with previous data \cite{Gausemel}, as seen in Table \ref{tab:gammaTable}. However, the $\beta$ feeding intensity of the 2277-keV level is nearly a factor of ten smaller than the value previously observed.

Direct $\beta$-feeding to the 1803-keV level was estimated based on the intensities of the $\gamma$-rays populating this state and the $\gamma$-ray intensities depopulating states below, connected through two transitions, a 41.0(2)-keV transition from the 1803-keV state to the 1762-keV state, and a 19.7(10)-keV transition from the 1762-keV state to the 1742-keV state. A direct $\beta$-feeding of 10(4)$\%$ to the 1803-keV state was observed, consistent with the previously reported value of 14(4)$\%$ $\beta$-feeding intensity observed previously \cite{Gausemel}. Table \ref{tab:betaFeeding-m2} also summarizes the $\beta$-feeding intensities and the log\textit{ft} values for the 2277-keV and 3993-keV states.

\begin{table}[h!]
\centering
 \caption{The $\beta$-feeding intensities and log\textit{ft} values, calculated for states in $^{129}$Sn, observed through the $\beta$-decay of the (23/2$^{-}$) $^{129}$In$^{m2}$ isomer and calculated with the weighted average half-life of 0.65(2) s. The values are calculated in this work are compared to those calculated by Gausemel \textit{et al.} \cite{Gausemel}. \label{tab:betaFeeding-m2}}
 \begin{tabular}{c|cc|cc}
    \hline \hline
\multirow{2}{*}{$E_{x}$ (keV)} & \multicolumn{2}{c}{$I_{\beta}$ (\%)} & \multicolumn{2}{c}{log\textit{ft}} \\ 
 \cline{2-5}  & This work & Ref. \cite{Gausemel} & This work & Ref. \cite{Gausemel} \\ \hline
1803(1)	&	10(4)	&	14(4)	&	5.92(18)	&	5.8(2)	\\
2277(1)	&	0.5(3)	&	8.0(12)	&	7.1(3)	&	5.9(1)	\\
3993(1)	&	89(4)	&	75(4)	&	4.31(2)	&	4.4(1)	\\
     \hline \hline
\end{tabular}
\end{table}

\subsection{Decay of (29/2$^{+}$) $^{129}\textrm{In}^{m3}$ \label{sub:m3State}}
Above the 1630-keV isomer in $^{129}$In, there is another isomer at 1911 keV with spin (29/2$^{+}$). This state has been shown to decay through a 281-keV internal transition to the 1630-keV $^{129}$In$^{m2}$ isomer. This transition lies very close in energy to two known transitions in the $^{129}$Sn nucleus, at 278.0(2)-keV and 280.4(2)-keV, which depopulate the 1047-keV and 1044-keV states, respectively, as seen in Figure \ref{fig:lvlScheme-gs}. 

The total relative intensity obtained in the addback singles $\gamma$-ray spectrum was 0.0805(4) for the triplet centered around 280-keV. This was inconsistent with the measured intensity of either the 278-keV and 280-keV transitions, observed in previous studies \cite{Gausemel}, and therefore the intensities of the transitions were measured by gating from below, on the 769-keV transition depopulating the 769-keV state for the transition at 278-keV, and the 728-keV transition that depopulates the 764-keV state for the 280-keV transition. 

\begin{figure}[h!]
\includegraphics[width=\columnwidth]{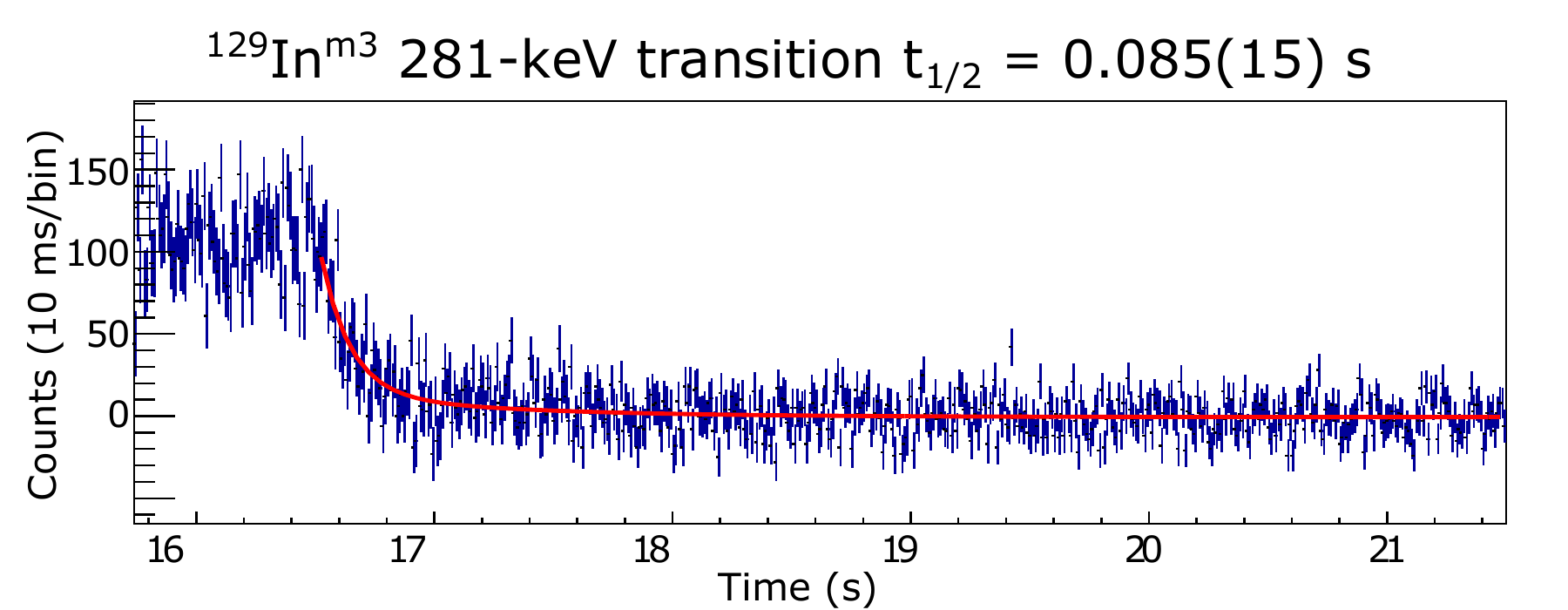}
\caption{A spectrum of total counts as a function of cycle time for the 281-keV transition. The fit, in red, returned a value of $t_{1/2}$ = 0.085(15) s, which is in good agreement with the half-life of the 1911-keV $^{129}$In$^{m3}$, quoted at 0.110(15) s \cite{NNDC-129In}. The fit returned a reduced $\chi^{2}$ of 1.1. The fit included a component associated with the $\beta$-decay of the $^{129}$In$^{gs}$ since the energy of the 281-keV internal transition in $^{129}$In is unresolved from two transitions in $^{129}$Sn at 278 and 280-keV (see the text for details).\label{fig:halfLife-280}}
\end{figure}

\begin{figure}[hb!]
\includegraphics[width=\columnwidth]{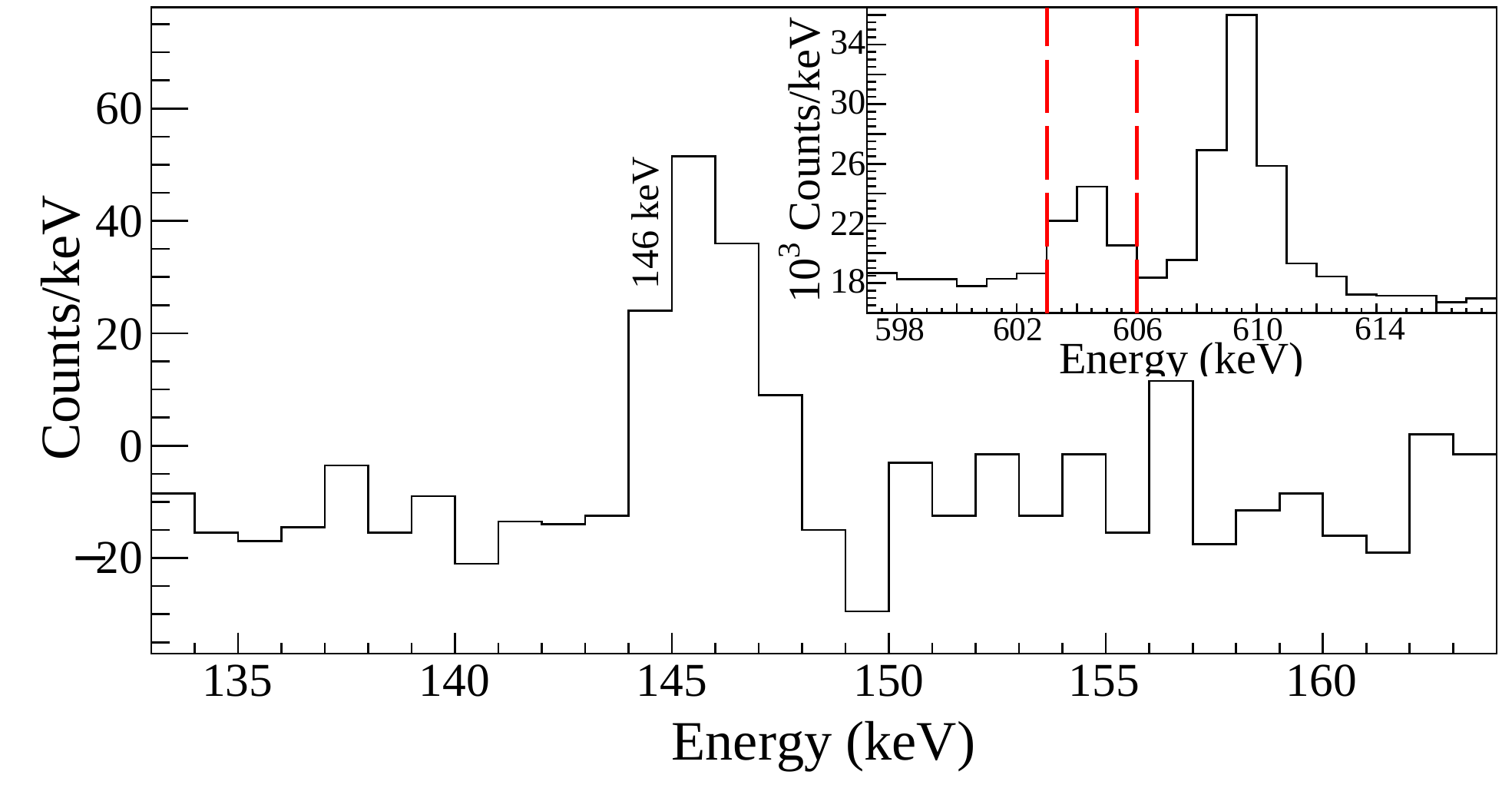}
\caption{A spectrum showing the 146-keV transition in the 604-keV gate. Note the red lines in the inset denote the placement of the gate.}\label{fig:604Gate}
\end{figure}

The relative intensity of the 278-keV transition was found to be 0.0080(10), consistent with previous measurements of both $\gamma$-ray intensity and the $\beta$-feeding of the 1047-keV state. Similarly, the relative intensity of the 280-keV transition in $^{129}$Sn was measured to be 0.0068(8), in agreement with the previously measured $\gamma$-ray intensity and $\beta$-feeding to the 1044-keV state.
The remaining intensity at 281-keV, amounting to 0.0657(11), must then be due to the internal transition of the $^{129}$In$^{m3}$. 

To confirm this assignment, a spectrum of total counts as a function of time, gated on the 281-keV transition was produced in the same manner as described in Sections \ref{subsub:halfLife-gs}, \ref{subsub:halfLife-m1} and \ref{subsub:halfLife-m2}. The fit, shown in Figure \ref{fig:halfLife-280}, returned a half-life value of 0.085(15) s, in good agreement with the 0.110(15) s half-life of the 1911-keV $^{129}$In$^{m3}$ state \cite{NNDC-129In} and much shorter than the half-lives of the other $\beta$-decaying states in $^{129}$In. This confirms that the presence of the excess intensity at 280 keV was due to the internal transition from $^{129}$In$^{m3}$ to $^{129}$In$^{m2}$.

\begin{figure}[h]
\includegraphics[width=0.9\columnwidth]{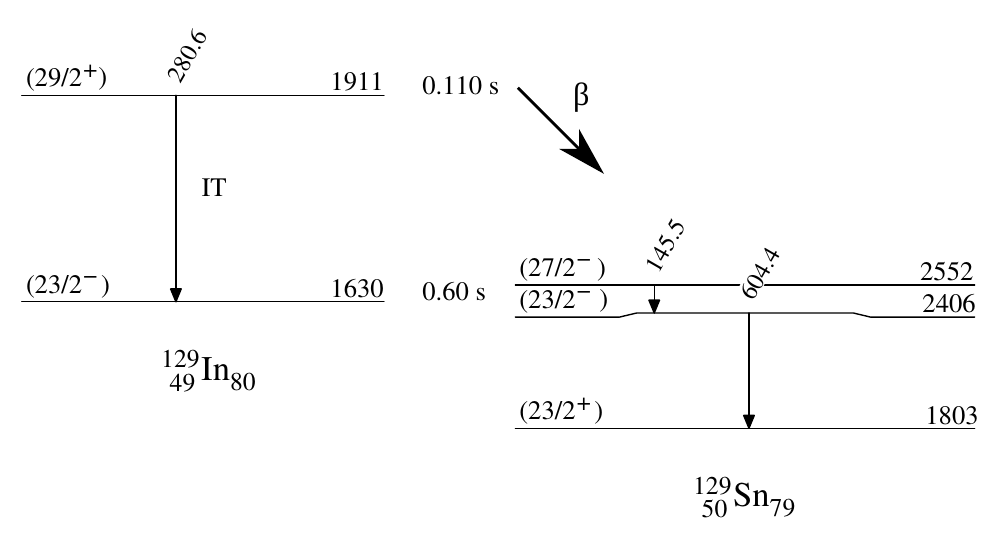}
\caption{Partial level scheme showing the decay of the (29/2$^{+}$) 1911-keV $^{129}$In$^{m3}$ isomer into the (27/2$^{-}$) 2552-keV state in $^{129}$Sn. The intensity of the 146-keV transition was obtained in coincidence with the 604-keV transition.\label{fig:lvlscheme-m3}}
\end{figure}

If this 1911-keV $^{129}$In$^{m3}$ isomer were to populate states in $^{129}$Sn through $\beta$-decay, these would have to be very high spin states. One such candidate is the 2552-keV state, with a proposed spin of (27/2$^{-}$). Lozeva \textit{et al.} \cite{Lozeva} observed a 146-keV transition from this state to the (23/2$^{-}$) 2406-keV state, having populated states in $^{129}$Sn through $^{238}$U fission and $^{136}$Xe fragmentation experiments. The present work observed the 146-keV transition in coincidence with the 604-keV transition from the 2406-keV state to the (23/2$^{+}$) 1803-keV state, also observed by Lozeva \textit{et al.} and shown in Figure \ref{fig:lvlscheme-m3}. Figure~\ref{fig:604Gate} shows the coincidence spectrum, produced by gating on the 604-keV transition. This gate clearly shows the presence of the coincidence with the 146-keV $\gamma$-ray.

An intensity balance calculation of the 2552-keV state implies $\beta$-feeding, since there is no known higher-lying state in $^{129}$In that could potentially populate this state. This feeding would amount to the portion represented by the intensity of the 146-keV transition, with respect to the sum of the intensities of the 146-keV and the 281-keV transitions, corrected for internal conversion.

Comparing the $\beta$-feeding of the 2552-keV state and the 281-keV internal transition yielded a $\beta$-branching ratio of 2.0(5)$\%$. The log\textit{ft} value for the $\beta$-decay of the (29/2$^{+}$) 1911-keV isomer in $^{129}$In to the (27/2$^{-}$) 2552-keV state in $^{129}$Sn is then 5.68(12), consistent with a first-forbidden transition from the (29/2$^{+}$) $^{129}$In$^{m3}$ state to the (27/2$^{-}$) state in  $^{129}$Sn. This log\textit{ft} value is calculated with a half-life of 0.10(1) s, the weighted average of the present result with the literature value. 
 The log\textit{ft} value in this transition is comparable to the 5.8 observed by Gausemel \textit{et al.} between the (23/2${-}$) 1630-keV isomer in $^{129}$In to the (23/2$^{+}$) 1803-keV state $^{129}$Sn, which is expected given the nearly pure $\pi(g_{9/2}^{-1}) \rightarrow \nu(h_{11/2}^{-1})$ $\beta$ transition \cite{Gausemel}.
This is the first time the 1911-keV isomer in $^{129}$In has been observed to $\beta$-decay to excited states of $^{129}$Sn. Figure \ref{fig:lvlscheme-m3} shows the partial level schemes of the states in both $^{129}$In and $^{129}$Sn involved in this decay.

\section{Conclusion\label{sec:conclusion}}
The present work reports new information observed through the $\beta$-decay of $^{129}$In to $^{129}$Sn. The half-lives of the ground state and isomeric states in  $^{129}$In have been confirmed, with the uncertainty in the half-life value of the $^{129}$In$^{m2}$ isomer improved. The level scheme of $^{129}$Sn has been greatly expanded, with nine new excited states and thirty-one new $\gamma$-ray transitions. Furthermore, this work observed, for the first time, the $\beta$-decay of the (29/2$^{+}$) 1911-keV $^{129}$In$^{m3}$ state.
More work is needed, in particular in the determination of the spins and parities of the states above the ground state, the 35-keV isomer and the 315-keV first excited state of $^{129}$Sn. This work provides more rigid constraints on the spins of a number of the excited states, but  further studies are needed in order to properly assign these values.
This new information on these two nuclei, lying close to doubly magic $^{132}_{\texttt{ }50}$Sn$_{82}$, provides important constraints and will guide future theoretical models in this region. 

\section*{Acknowledgements\label{sec:acknow}}
The authors would like to thank the GRSI collaboration at TRIUMF for their aid during the course of this work. This work was supported, in part, by the Natural Sciences and Engineering Research Council of Canada (NSERC). C.E.S. acknowledges support from the Canada Research Chairs program. Phase I of the GRIFFIN spectrometer was funded by the Canadian Foundation of Innovation (CFI), TRIUMF and the University of Guelph. TRIUMF receives funding from the Canadian Federal Government via a contribution agreement with NRC.

\bibliographystyle{apsrev4-1}
\bibliography{129InPaper}

\end{document}